%% file: v3.tex
\documentstyle[12pt,epsfig,epsf,feynarts]{article}
\def\ga{\mathrel{\raise.3ex\hbox{$>$\kern-.75em\lower1ex\hbox{$\sim$}}}}
\def\la{\mathrel{\raise.3ex\hbox{$<$\kern-.75em\lower1ex\hbox{$\sim$}}}}

\begin{document}
\begin{flushright}
LPHEP-04-03\\
September 2004 \\
\end{flushright}

\vspace*{.8cm}
\begin{center}
{\large{{\bf 
Higgs bosons decay into bottom-strange\\ 
in two Higgs Doublets Models }}}

\vspace{1.1cm}
Abdesslam Arhrib \\
D\'epartement de Math\'ematiques, Facult\'e des Sciences et Techniques\\
B.P 416 Tanger, Morocco.\\
and\\
LPHEA, D\'epartement de Physique, Facult\'e des Sciences-Semlalia,\\
B.P. 2390 Marrakech, Morocco.\\
\end{center}

\begin{abstract}
We analyze the decays $\{h^0,H^0,A^0\}\to \bar{s}b$
within two Higgs Doublet Models with Natural Flavor Conservation 
(2HDM) type I and II.
It is found that the Higgs bosons decay
into bottom-strange can lead 
to a branching ratio
in the range $10^{-5}\to 10^{-3}$ for small $\tan\beta \approx
0.1\to 0.5$ and rather light charged Higgs in the 2HDM type I. 
When $\tan\beta \ga 1$, one can easily
reach a branching ratio of the order $10^{-5}$. 
In 2HDM type II, without imposing $b\to s\gamma$ constraint, 
the situation is the same as in 2HDM type I.
If $b\to s\gamma$ constraint 
on charged Higgs mass ($M_{H\pm}\geq 350$ GeV) is imposed, 
we obtain $Br(h^0 \to \bar{s}b)$ in the range $10^{-5}$--$10^{-6}$.
A comparison between the rates of 
$h^0\to \bar{s}b$ and $h^0\to \gamma \gamma$
is made. It is found that in the fermiophobic scenario,
$h^0\to \gamma \gamma$ is still the dominant decay mode. 
\end{abstract}

\newpage
\section{Introduction}
\label{sec:1}
One of the goals of the next generation of 
high energy colliders, such as the large hadron
collider LHC \cite{LHC} or the linear collider LC \cite{TESLA} 
or muon colliders, is to probe top Flavor-Changing Neutral 
Couplings `top FCNC' as well as the Higgs 
Flavor-Changing Neutral Couplings `Higgs FCNC'. 
FCNC of heavy quarks have been 
intensively studied both from the theoretical and experimental point of view. 
Such processes are being well established in the Standard Model (SM)
and are excellent probes for the presence 
of new physics effects such as Supersymmetry, 
extended Higgs sector and extra fermions families.
 
Within the SM, with one Higgs doublet, the FCNC $Z \bar{t} c$ 
vanishes at tree-level
by the GIM mechanism, while the
$\gamma \bar{t}c$ and $g\bar{t}c$ couplings are zero as a
consequence of  the unbroken $SU(3)_c \times {U(1)}_{\mbox{em}}$ 
gauge symmetry. The Higgs FCNC $H\bar{t}c$ and $H\bar{s}b$ 
couplings also vanish 
due to the existence of only one Higgs doublet.
Both top FCNC and Higgs FCNC are generated at one loop level
by charged current exchange, but they are very suppressed by 
the GIM mechanism. The calculation of the branching ratios for top decays
yields the SM predictions \cite{sm1}, \cite{sm2}:
\begin{eqnarray}
& & \mathrm{Br}(t \to Zc) = 1.3 \times 10^{-13},
\mathrm{Br}(t \to \gamma c) = 4.3 \times 10^{-13},
\mathrm{Br}(t \to gc) =  3.8 \times 10^{-11},\nonumber\\
& & \mathrm{Br}(t \to Hc) = 5.6 \to 3.2 \times 10^{-14}\qquad   
\mathrm{for}\qquad M_H=115 \to 130\ \ \mathrm{GeV}.\label{eq2}
\end{eqnarray}
While for Higgs FCNC, calculation within SM leads to:
\begin{eqnarray}
& & \mathrm{Br}(H \to \bar{s}b) \approx  10^{-7}\ \  
(\mathrm{resp}\  10^{-9})\ \ m_H=100\ \ (\mathrm{resp}\ 200) \ 
\mathrm{GeV}\nonumber \\ 
& & \mathrm{Br}(H \to \bar{t}c) \approx 1.5\times 10^{-16} \ \
(\mathrm{resp}\ 3\times 10^{-13})\ \ m_H=200\ \  (\mathrm{resp}\ 500) \ 
\mathrm{GeV}\label{eq3} 
\end{eqnarray}

Many SM extensions predict that 
these  top and Higgs FCNC can be orders of magnitude 
larger than their SM values (see \cite{review} for an overview). 
For the Higgs FCNC, an important class of models where 
Higgs FCNC appear at tree level are the so called Two Higgs Doublet
Model without Natural Flavor Conservation (NFC) 2HDM-III 
\cite{sher,2hdm2,2hdm1,soni}. In this class of models,
the branching ratio of $h\to \bar{t}c$ can be larger than 10\% 
in some cases \cite{2hdm2}.
In the framework of 2HDM with NFC type I and II,
top and Higgs FCNC have been studied in \cite{2hdm44,2hdm4}. It was shown
that in 2HDM-II the $Br(\Phi\to \bar{t}c)$, $\Phi=h^0$ or $H^0$, 
may reach  $10^{-5}$ for CP-even states \cite{2hdm4}. 
This rate is almost eight orders of magnitude larger than the SM one.\\
Top and Higgs FCNC couplings have been addressed also in supersymmetry 
\cite{maria1,maria2,bdgs,susy3,susy1}. 
In those studies it has been shown 
that $Br(h^0\to \bar{s}b)$ can be in the range of $10^{-4}$-$10^{-3}$.
This rate originates mainly from the flavor violation interactions 
mediated by the gluino \cite{maria1,bdgs}.
In case of  MSSM with R parity conservation,
the top FCNC
coupling $t\to c h^0$, can reach $10^{-4}$ branching ratio \cite{susy3}
in case of flavour violation induced by gluino.

Hence, Higgs and top FCNC offer a good place to search for new
physics, which may manifest itself if those couplings are observed in
future experiments such as LHC or LC \cite{LHC,TESLA}. 
Therefore, models which can enhance those FCNC couplings are welcome.

The aim of this paper is to study  Higgs FCNC 
couplings such as $\Phi \to \bar{s}b$, $\Phi=h^0, H^0, A^0$, 
in the framework of NFC two Higgs Doublet  Models type I and II. 
It is found that the branching ratios of
$Br(\Phi\to \bar{s}b)$, $\Phi=h^0, H^0, A^0$, can be 
greater than $\ga 10^{-5}$ in quite a substantial region of the 2HDM
parameters space. $Br(\Phi\to \bar{t}c)$ requires large $\tan\beta$ and
light charged Higgs \cite{2hdm4} while $Br(\Phi\to \bar{s}b)$ requires 
rather small $\tan\beta$ together with light charged Higgs 
and large soft breaking term $\lambda_5$.\\
We would like to mention here that due to the isolated top quark signature,
Higgs FCNC $\Phi\to \bar{t}c$ event may be easy to search for experimentally.
However, it is very difficult to isolate
Higgs FCNC $\Phi\to \bar{s}b$ events from the background. 

The paper is organized as follows. 
In the next section, the 2HDM is introduced. Relevant couplings are
given, theoretical and experimental constraints 
on 2HDM parameters are discussed.
In the third section, we will study the effects 
of 2HDM on $Br(\Phi\to \bar{s}b)$ which are evaluated 
in 2HDM-I and 2HDM-II. A comparison between 
$Br(h^0\to \bar{s}b)$ and $Br(h^0\to \gamma \gamma)$ is also 
discussed. Our conclusion is given in section 4. 

\section{The 2HDM}
Two Higgs Doublet Models (2HDM) are formed by adding an extra complex
$SU(2)_L\otimes U(1)_Y$ scalar doublet to the SM Lagrangian. 
Motivations for such a structure include CP--violation in the Higgs 
sector and the fact that some models of
dynamical electroweak symmetry breaking 
yield the 2HDM as their low-energy effective theory \cite{dewsb}.

\begin{figure}[t!]
\begin{center}
\vspace{-2.3cm}
\input{hsbb.tex}
\vspace{-8.7cm}
\caption{Generic contribution to $\Phi \to f_1 f_2$ in SM $d_1\to d_{10}$,
in 2HDM $d_{11}\to d_{18}$}
\label{hsbb}
\end{center}
\end{figure}
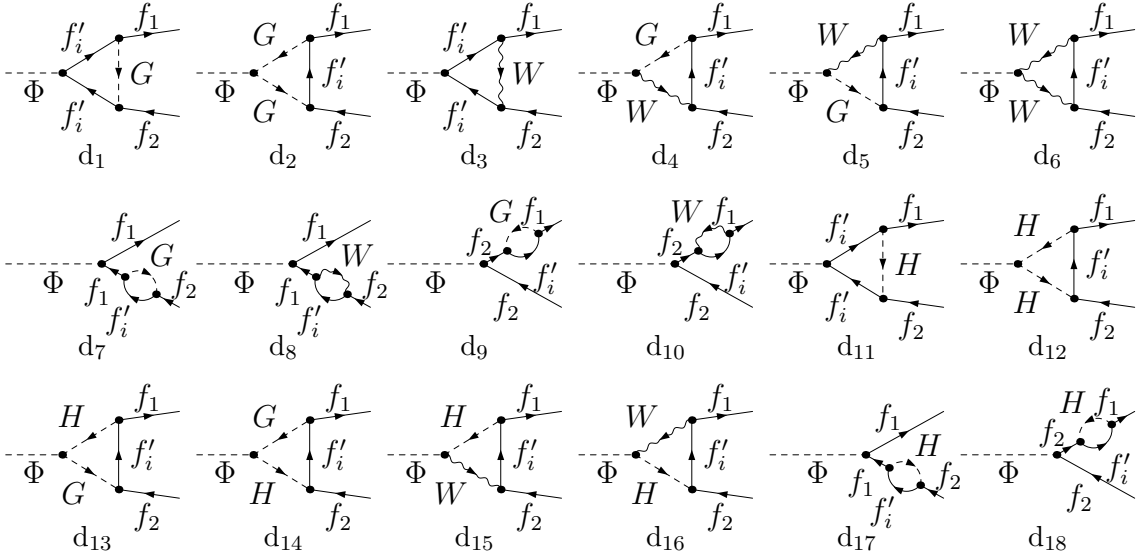

The most general 2HDM scalar potential which is both 
$SU(2)_L\otimes U(1)_Y$ and CP invariant is given by \cite{Gun}:
\begin{eqnarray}
 V(\Phi_{1}, \Phi_{2})& & =  \lambda_{1} ( |\Phi_{1}|^2-v_{1}^2)^2
+\lambda_{2} (|\Phi_{2}|^2-v_{2}^2)^2+
\lambda_{3}((|\Phi_{1}|^2-v_{1}^2)+(|\Phi_{2}|^2-v_{2}^2))^2 
+\nonumber\\ [0.2cm]
&  & \lambda_{4}(|\Phi_{1}|^2 |\Phi_{2}|^2 - |\Phi_{1}^+\Phi_{2}|^2  )+
\lambda_{5} (\Re(\Phi^+_{1}\Phi_{2})
-v_{1}v_{2})^2+ \lambda_{6} [\Im(\Phi^+_{1}\Phi_{2})]^2 
\label{higgspot}
\end{eqnarray}
where $\Phi_1$ and $\Phi_2$ have weak hypercharge Y=1, $v_1$ and
$v_2$ are respectively the vacuum
expectation values of $\Phi_1$ and $\Phi_2$ and the $\lambda_i$
are real--valued parameters. 
Note that this potential violates the discrete symmetry
$\Phi_i\to -\Phi_i$ softly by the dimension two term
$\lambda_5 \Re(\Phi^+_{1}\Phi_{2})$.
The above scalar potential has 8 independent parameters
$(\lambda_i)_{i=1,...,6}$, $v_1$ and $v_2$.
After electroweak symmetry breaking, the combination $v_1^2 + v_2^2$ 
is thus fixed by the electroweak 
scale through $v_1^2 + v_2^2=(2\sqrt{2} G_F)^{-1}$.
We are left then with 7 independent parameters.\\
Meanwhile,  three of the eight degrees of freedom 
of the two Higgs doublets correspond to 
the 3 Goldstone bosons ($G^\pm$, $G^0$) and  
the remaining five become physical Higgs bosons: 
$H^0$, $h^0$ (CP--even), $A^0$ (CP--odd)
and $H^\pm$. Their masses are obtained as usual
by diagonalizing the mass matrix. 
The presence of charged Higgs bosons will give new contributions
to the one--loop induced FCNC couplings, as shown in Fig.~(\ref{hsbb}) 
$d_{11}\to d_{18}$.

It is possible to write the $\lambda_i$ in terms of 
physical scalar masses, $\tan\beta$, $\alpha$ and $\lambda_5$ 
(see \cite{AA} for details). We
are then free to take as 7 independent parameters 
$(\lambda_i)_{i=1,\ldots , 6}$ and $\tan\beta$
or equivalently the four scalar masses, $\tan\beta$, $\alpha$
and one of the $\lambda_i$. In what 
follows we will take $\lambda_5$ as a free parameter as well as the
physical masses and mixing.

We list hereafter the Feynman rules in the general 2HDM 
for the trilinear scalar couplings relevant for our study. 
They are written in terms of 
the physical masses, $\alpha$, $\beta$ and the soft 
breaking term $\lambda_5$:

\begin{eqnarray}
 {H^0H^+H^-}= & &\frac{-ig}{M_W \sin 2\beta } (
 M_{H^0}^2 (\cos^3\beta \sin\alpha +\sin^3\beta \cos\alpha)+
M_{H^{\pm}}^2\sin{2\beta} \cos({\beta-\alpha})\nonumber \\ & & -
\sin({\beta+\alpha})\lambda_5 v^2 )\label{scalar1}  \\
{H^0H^+G^-}  =& &  \frac{ig}{2 M_W} \sin({\beta-\alpha}) 
(M_{H^0}^2-M_{H^{\pm}}^2)\label{scalar2} \\
{h^0H^+H^-}  = && \frac{- ig}{M_W \sin 2\beta} ( 
 M_{h^0}^2(\cos{\alpha}\cos^3{\beta}-
\sin{\alpha}\sin^3{\beta})
 +M_{H^{\pm}}^2\sin{2\beta}\sin({\beta-\alpha})
\nonumber \\ && -
\cos({\beta+\alpha}){\lambda_5} v^2)\label{scalar3}   \\
{h^0H^+G^-} =&&\frac{-ig}{2 M_W}  \cos({\beta-\alpha}) 
(M_{h^0}^2-M_{H^{\pm}}^2) \label{scalar4}  \\
 {A^0H^+G^-}  = & & \frac{-g}{2 M_W}  
(M_{H^{\pm}}^2 - M_A^2)\qquad , \ \ v^2 = \frac{2M_W^2}{g^2} \label{scalar5}
\end{eqnarray}

We need also the couplings of scalar boson to a pair of fermions
both in 2HDM-I and 2HDM-II. In those couplings, the relevant terms 
are as follows:
\begin{eqnarray}
& & h^0\bar{t}t \propto M_t\frac{\cos\alpha}{\sin\beta} \ \ \ , \ \ \ 
H^0\bar{t}t \propto M_t \frac{\sin\alpha}{\sin\beta}\ \ \ , \ \ \ 
A^0\bar{t}t \propto \frac{M_t}{\tan\beta} \ \ {\rm 2HDM-I , II}
\label{coupl1}\\
& & h^0\bar{b}b \propto M_b\frac{\cos\alpha}{\sin\beta} \ \ \ , \ \ \ 
H^0\bar{b}b \propto M_b \frac{\sin\alpha}{\sin\beta}\ \ \ , \ \ \ 
A^0\bar{b}b \propto \frac{M_b}{\tan\beta}\ \ {\rm 2HDM-I }\label{coupl2}\\
& & h^0\bar{b}b \propto M_b\frac{\sin\alpha}{\cos\beta} \ \ \ , \ \ \ 
H^0\bar{b}b \propto M_b \frac{\cos\alpha}{\cos\beta}\ \ \ , \ \ \ 
A^0\bar{b}b \propto M_b\tan\beta \ \ {\rm 2HDM-II }\label{coupl3}\\
& & (H^-\bar{b}t)_L \propto \frac{M_b}{\tan\beta} \ \ \ , \ \ \ 
(H^-\bar{b}t)_R \propto \frac{M_t}{\tan\beta}  \ \ {\rm 2HDM-I } 
\label{coupl4}\\
& & (H^-\bar{b}t)_L \propto {M_b}{\tan\beta} \ \ \ , \ \ \ 
(H^-\bar{b}t)_R \propto \frac{M_t}{\tan\beta}  \ \ {\rm 2HDM-II } 
\label{coupl5}
\end{eqnarray}

In this analysis, we take into account the following 
constraints when the independent parameters are varied.
From the theoretical point of view:\\
$i)$ The contributions to the $\delta\rho$ parameter from the Higgs
scalars \cite{hollik} should not exceed the current limits from precision 
measurements \cite{PDG}: $|\delta\rho|\la 0.001$.
We stress in passing that the extra contribution to 
$\delta\rho$ constraint \cite{hollik} vanish when 
we take $M_{H^\pm}=M_A$ ($\lambda_4=\lambda_6$).
Under this constraint the 2HDM scalar 
potential is $O(4)$ symmetric \cite{negative}. In this case
$(H^+,A^0,H^-)$ form a triplet under the residual global $SU(2)$ of
the Higgs potential. It is this residual symmetry which ensures that 
$\rho$ is equal to unity at tree level. One conclude then that large
splitting between $M_{H^\pm}$ and $M_A$ could violate 
$|\delta\rho|\la 0.001$ constraint.
\\
\noindent
$ii)$ From the requirement of perturbativity for the
top and bottom Yukawa couplings \cite{berger}, $\tan\beta$ is 
constrained to lie in the range $0.1\leq \tan\beta \leq 70$. \\
$iii)$ It has been shown in~\cite{bsg} that 
for models of the type 2HDM-II, data on $b\to s \gamma$ 
imposes a lower limit of
$M_{H^\pm} \ge 350$\,GeV.
In type I 2HDM, there is no such a constraint
on charged Higgs mass \cite{bsg}. 
In our numerical analysis we will ignore this 
constraint in order to localize regions in the 2HDM parameters
space where the branching ratios are sizeable.\\
$iv)$ Unitarity and perturbativity constraints on scalar parameters:\\
It is well known that the unitarity bounds coming from a tree-level 
analysis~\cite{abdesunit,kan} put severe constraints
on all scalar trilinear and quartic couplings. 
The tree level unitarity bounds are derived with 
the help of the equivalence theorem, which itself is a 
high-energy approximation where it is assumed that the 
energy scale is much larger than the $Z^0$ and $W^\pm$ 
gauge-boson masses.  We will use, instead
of unitarity constraints, the perturbativity constraints
by assuming that all $\lambda_i$ satisfy:
\begin{eqnarray}
|\lambda_i| \leq 4 \pi .\label{pert}
\end{eqnarray}
Those perturbative constraints on the 
$\lambda_i$ allow us to investigate a larger parameter space
than the one allowed by unitarity constraints.\\
We would like to mention also that when performing the scan over the 
2HDM parameters space, we realize that 
for some points the widths $\Gamma_{\Phi}$ of the scalar particles 
become bigger than their corresponding masses:
$\Gamma_{\Phi} \geq M_{\Phi}$ ($\Phi=h^0,H^0,A^0,H^\pm$).
This happens both when we impose tree level unitarity constraints
and/or perturbativity constraints. The width becomes large
specially when the pure scalar decays like $H^0\to h^0h^0$, 
$H^0\to H^+H^-$, $h^0\to H^+H^-$, $H^0\to A^0A^0$
and $h^0\to A^0A^0$ are open.
We find it is natural to add to the
above constraints the requirement that the width of the scalar
particles remains smaller than the mass 
of the corresponding particles:
\begin{eqnarray}
\Gamma_{\Phi}< M_{\Phi} \label{wid}
\end{eqnarray}

From the experimental point of view,
the combined null--searches from all four CERN LEP collaborations derive the 
lower limit $M_{H^{\pm}}\ge 78.6$ GeV $(95\%\, CL)$, a limit
which applies to all models in which Br($H^{\pm}\to \tau\nu_{\tau}$)+
Br($H^{\pm}\to c\bar{s}$)=1. For the neutral Higgs bosons,
OPAL collaboration has put a limit on 
$h^0$ and $A^0$ masses  
of the 2HDM. They conclude that the regions
$1\la M_h \la 44$ GeV and $12\la M_A \la 56$ GeV 
are excluded at 95\% CL
independent of $\alpha$ and $\tan\beta$ \cite{opal}.
For simplicity we will assume that all scalar particles 
masses are $\ga 90$ GeV.

\section{Higgs FCNC in 2HDM}

\subsection{Higgs FCNC in SM}
Before presenting our results in 2HDM, we would like to 
give the Branching ratio of $H\to {\bar{t}}c$ 
and $H\to {\bar{s}}b$ in the SM.
To our best knowledge, the first calculation for $Br(H\to {\bar{s}}b)$
has been carried out in \cite{sonism}. However, in  \cite{sonism},
numerical results have been given only for a very light Higgs boson
$M_H=9$ GeV. Recently a new estimation,
using dimensional analysis and power counting,
has appeared both for $Br(H\to {\bar{s}}b)$
 \cite{bdgs} and $Br(H\to {\bar{t}}c)$ \cite{2hdm4}.
We refer the reader to \cite{2hdm4,bdgs} for more details
on those estimations.
Here we present exact result based on diagrammatic calculations
both for $Br(H\to {\bar{s}}b)$ and $Br(H\to {\bar{t}}c)$.
We give numerical results for the width as well as for the
branching ratio.\\
The Feynman diagrams contributing to those process in SM are depicted
in Fig~.(\ref{hsbb}) d$_1 \to$ d$_{10}$.
In the case of $H\to {\bar{t}}c$, in Fig.~(\ref{hsbb})
$(f_1,f_2)=(t,c)$ and $f_i^\prime=d,s,b$, while 
for $H\to {\bar{s}}b$
$(f_1,f_2)$ is $(b,s)$ and $f_i^\prime=u,c,t$.
The full loop calculation presented here is done with the help of 
FormCalc \cite{FA2}. FF and LoopTools packages \cite{FF} 
are used in numerical analysis. The numerical results shown
in eqs.~(\ref{eq2},\ref{eq3}) is derived by FormCalc \cite{FA2}.\\
\begin{figure}[t!]
\smallskip\smallskip 
\vskip-3.8cm
\centerline{{
\epsfxsize2.8 in 
\epsffile{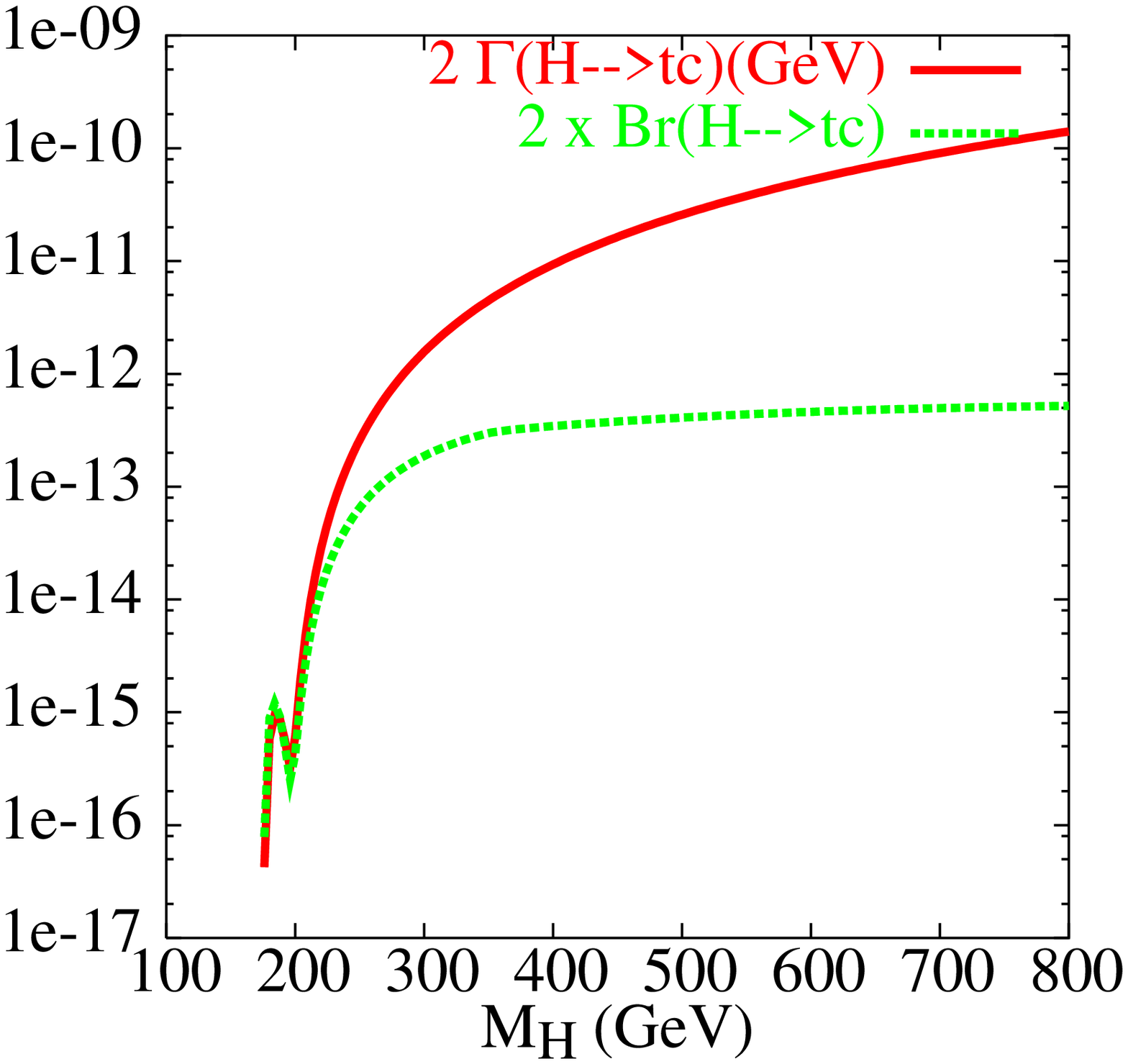}}  
\hskip0.4cm
\epsfxsize2.88 in 
\epsffile{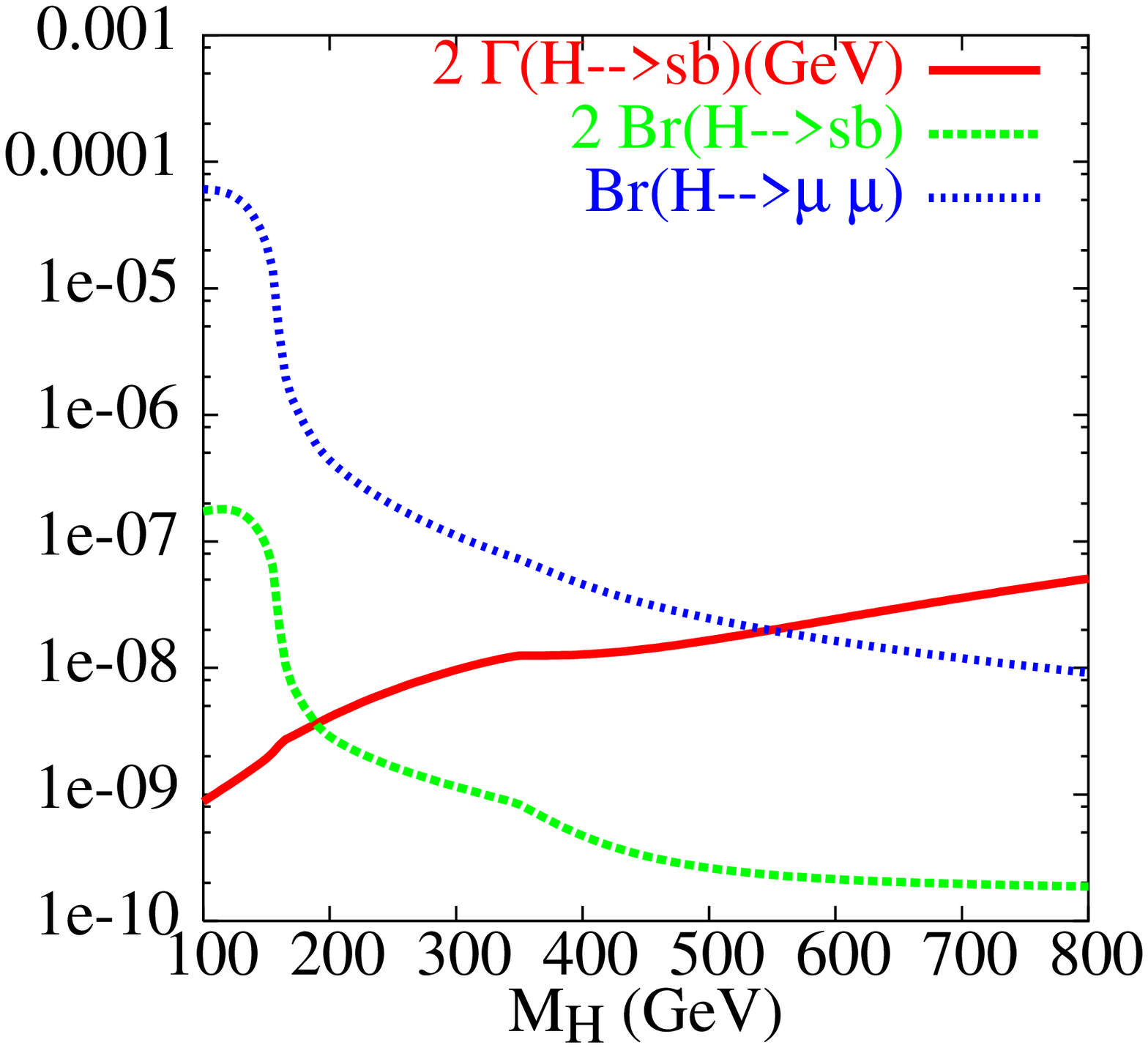} }
\smallskip\smallskip
\caption{SM width and Branching ratio for 
$H\to \bar{t}c$ (left) and 
$H\to \bar{s}b$ (right) 
as a function of Higgs mass.}
\label{fig2}
\end{figure}
In the SM, as expected, the branching ratio of $H\to \bar{t}c$ and
$H\to \bar{s}b$ 
are very suppressed due to GIM mechanism.
The branching ratio is very small in both cases for 
higher Higgs mass $M_H\geq 2 M_Z$ where 
$H\to W^+W^-$ and $H\to Z^0Z^0$ are open.

Both in SM and 2HDM, the decay widths $\Gamma_{\Phi}^{\rm{SM}}$ 
and $\Gamma_{\Phi}^{\rm{2HDM}}$ of  scalar particles: 
$\Phi=H^{\rm{SM}}$, $h^0$, $H^0$, $A^0$ and
$H^\pm$ are computed at tree level as follows: 
\begin{eqnarray}
\Gamma_{\Phi}^{\rm{SM}}=\sum_{f}\Gamma(\Phi\to f\bar{f}) + 
\Gamma(\Phi\to VV)  \nonumber\\
\Gamma_{\Phi}^{\rm{2HDM}}=\sum_{f}\Gamma(\Phi\to f\bar{f}) + 
\Gamma(\Phi\to VV)  +  \Gamma(\Phi\to V H_i)
+\Gamma(\Phi\to H_i H_j)\label{widd}
\end{eqnarray}
QCD corrections 
to $\Phi \to f\bar{f}$ and  $\Phi \to \{ g g, \gamma \gamma, \gamma Z,
V^*V^*, VV^*, V^*H_i\}$ decays are not included in the widths.
The decay widths of the Higgs bosons are taken from
\cite{AKZ}.

For a Higgs mass heavier than 250 GeV, we get branching ratio
of the order $10^{-14} \to 10^{-12}$ (resp $10^{-10} \to 10^{-9}$) for 
$H\to {\bar{t}}c$  (resp $H\to {\bar{s}}b$).
\\
In the case of $H\to {\bar{s}}b$, the branching ratio
is enhanced for Higgs boson mass of the order $M_H\approx 100 \to 120$ GeV
where the width of the Higgs is very narrow.
We have plotted in Fig.~(\ref{fig2}) both the decay width and the branching
ratios of $H\to {\bar{t}}c$ (left plot) 
and $H\to {\bar{s}}b$ (right plot) as well as the branching ratio 
of $H\to \mu^+\mu^-$.  As it can be seen from the right plot
$Br(H\to \bar{s}b)$ is two orders of magnitude smaller 
than $Br(H\to \mu^+\mu^-)$.\\
Since the decay width of $H\to {\bar{t}}c$ is very suppressed,
the threshold for $t\bar{t}$ production is  absent in Fig.~2 (left). 
The situation is slightly different for $H\to \bar{s}b$
where the decay width of $H\to \bar{s}b$ is about 6 order of magnitude 
larger than decay width of $H\to \bar{t}c$. From the right plot of
Fig.~(\ref{fig2}) one can see that the Br of $H\to \bar{s}b$ 
is smaller once the $t\bar{t}$ threshold has been passed.

\subsection{$h^0 \to {\bar{s}}b$}

Turning now to the 2HDM Higgs bosons FCNC couplings $\Phi \to {\bar{s}}b$, 
$\Phi=h^0,H^0,A^0$. The Feynman diagrams are depicted in
Fig.~(\ref{hsbb}). The amplitude 
is sensitive to the $\Phi H^+ H^-$ and 
$\Phi H^\pm G^\mp $ couplings through diagrams $d_{12,13,14}$
as well as to the $\Phi t\bar{t}$ and 
$(H^- \bar{b}t)_{L,R}$ couplings through diagrams $d_{11,12,13,14}$. 
In 2HDM, it is expected that the dominant contribution to the 
amplitude of $\Phi^0\to \bar{s}b$ comes from diagram $d_{12}$.
The amplitude of $d_{12}$
is proportional to the trilinear Higgs coupling $\Phi^0H^+H^-$
and is given by ($\Phi=h^0, H^0$):
\begin{eqnarray}
M_{d_{12}}=\Phi^0H^+H^- \frac{\alpha V_{ts}}{8 \pi}
\frac{M_t^2}{\tan^2\beta}\frac{M_b}{8 M_W^2 s_W^2} 
[(1 + \tan\beta Y_b) C_0 + C_1 + C_2]\bar{v}(M_s)\frac{1+\gamma_5}{2}u(M_b)
\end{eqnarray}
where we have neglected the strange quark mass.
In the conventions of \cite{FA2},
the arguments of the Passarino-Veltman functions $C_i$ are
$\{M_b^2, M_s^2, M_\Phi^2, M_{H\pm}^2, M_t^2, M_{H\pm}^2\}$.
The Yukawa coupling $Y_b$ of the bottom  is model dependant
and is given by $Y_b=-1/\tan\beta$ (resp $Y_b=\tan\beta$) 
for 2HDM-I (resp 2HDM-II).\\
In 2HDM-I, $1 + \tan\beta Y_b=0$, the amplitude of $d_{12}$ 
is enhanced by $\frac{M_t^2}{\tan^2\beta}$
factor for small $\tan\beta$ as well as by the 
trilinear coupling $\Phi^0H^+H^-$.

The diagram $d_{11}$ is sensitive to the coupling $\Phi^0\bar{t}t$.
It is clear from  equation (\ref{coupl1}) that the top effect 
is enhanced for small $\tan\beta$ in the case of CP-odd $A^0$ boson.
While in the case of CP-even $H^0$ and $h^0$, the couplings are enhanced
 at small $\tan\beta$ and large $\sin\alpha$ (resp large $\cos\alpha$)  
for $H^0$ (resp $h^0$). 
Consequently, our numerics are presented for
small $\tan\beta=0.3$, $\sin\alpha=0.1$ for $h^0 \to {\bar{s}}b$
and $\sin\alpha=0.95$ for $H^0 \to {\bar{s}}b$.\\
We also give other numerical results for specific 2HDM parameters
where $Br(h^0\to \bar{s}b)$ and $Br(H^0\to \bar{s}b)$
get their maximum values without violating $\delta\rho$ and
perturbativity constraints.
\begin{figure}[t!]
\smallskip\smallskip 
\vskip-.1cm
\centerline{{
\epsfxsize2.83 in 
\epsffile{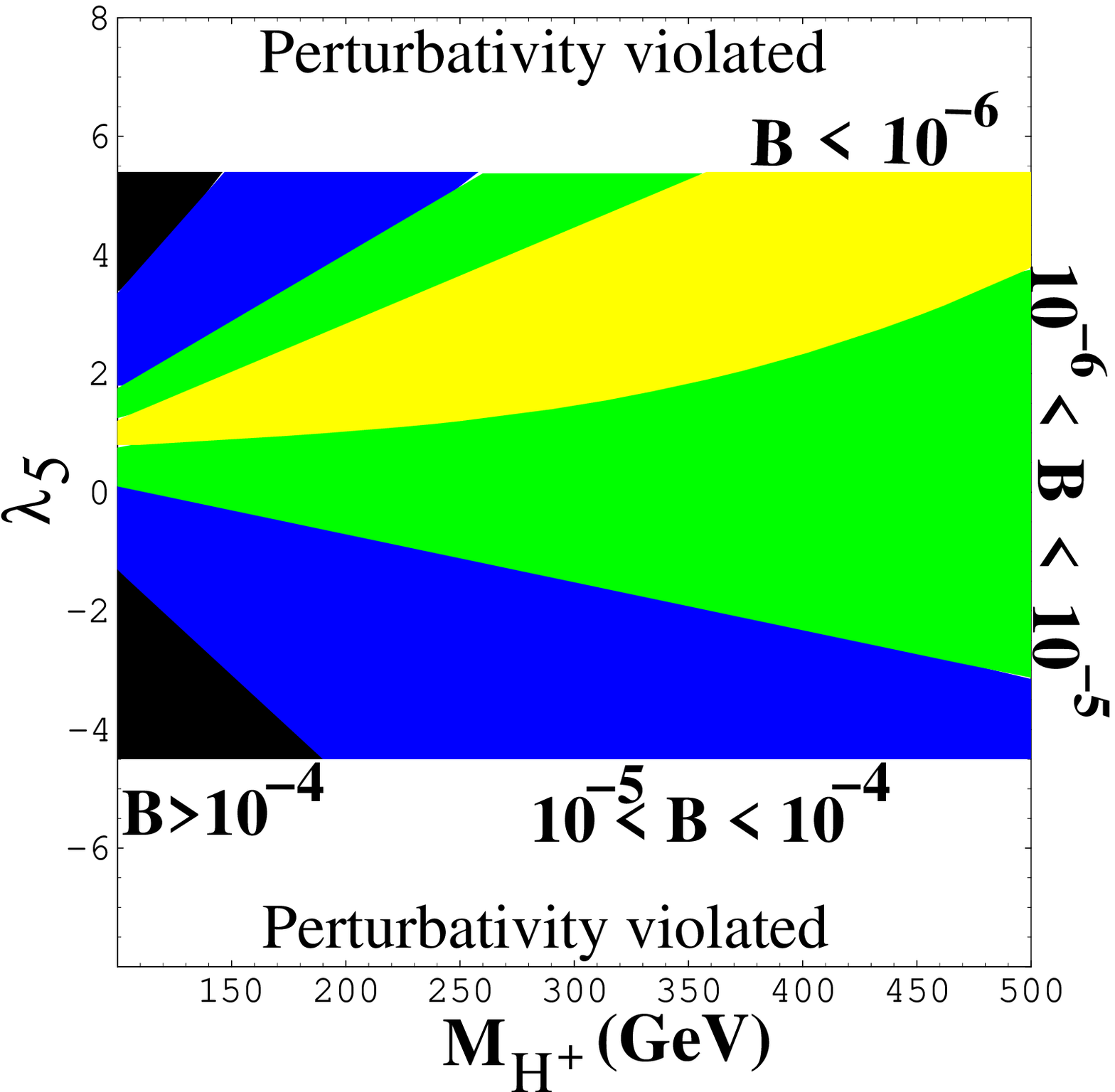}}  \hskip0.4cm
\epsfxsize2.8 in 
\epsffile{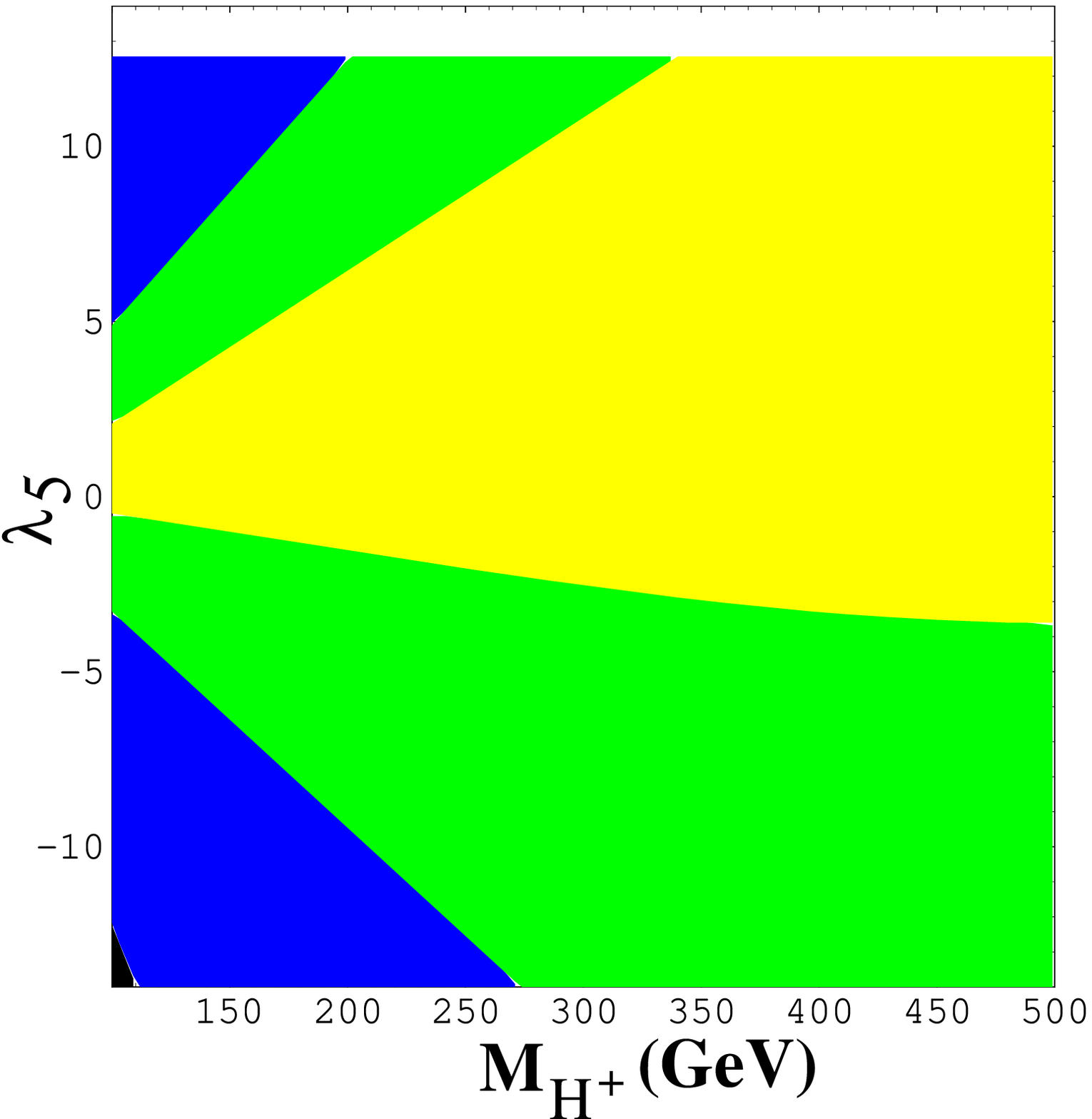}}
\smallskip\smallskip
\caption{Contours for $2\times Br(h^0\to  \bar{s}b)$ in 
2HDM-II $\tan\beta=0.3$ (left),
$\tan\beta=1.5$ (right) in the ($M_{H\pm}$, $\lambda_5$) plane with
$M_h=110$ GeV, $M_H=180$ GeV, $\sin\alpha=0.1$ and $M_{A^0}=M_{H\pm}$}
\label{fig9}
\end{figure}

We show in Fig.~(\ref{fig9}) 
contour plots for $Br(h^0\to  \bar{s}b)$ in 
2HDM-II $\tan\beta=0.3$ (left) and 
$\tan\beta=1.5$ GeV (right) in the ($M_{H\pm}$, $\lambda_5$) plane.
$\lambda_5$ is varied in the perturbative range $|\lambda_5|< 4 \pi$.
The other inputs are $M_h=110$ GeV, $M_H=180$ GeV, $\sin\alpha=0.1$ 
and $M_{A^0}=M_{H\pm}$. The width $\Gamma_{h^0}$ is computed at
tree level according to eq.~(\ref{widd}). 
Since the mass of $h^0$ is taken at 110 GeV, 
only light fermions contribute to $\Gamma_{h^0}$ and so the width is
very narrow and is of the order $57\times 10^{-4}$ (resp $83\times 10^{-5}$
GeV) at $\tan\beta=0.3$ (resp $\tan\beta=1.5$). Such narrow width
could enhance the branching ratio $Br(h^0\to  \bar{s}b)$.
We would like to mention first that for this set of parameters, the 
perturbativity of scalar quartic couplings $\lambda_i$ is violated around 
$\lambda_5\ga 5.5$. We get $|\lambda_1|>4 \pi$ for
$\tan\beta=0.3$, while for $\tan\beta=1.5$ there is no such bound.

\begin{figure}[t!]
\smallskip\smallskip 
\vskip-3.8cm
\centerline{{
\epsfxsize2.8 in 
\epsffile{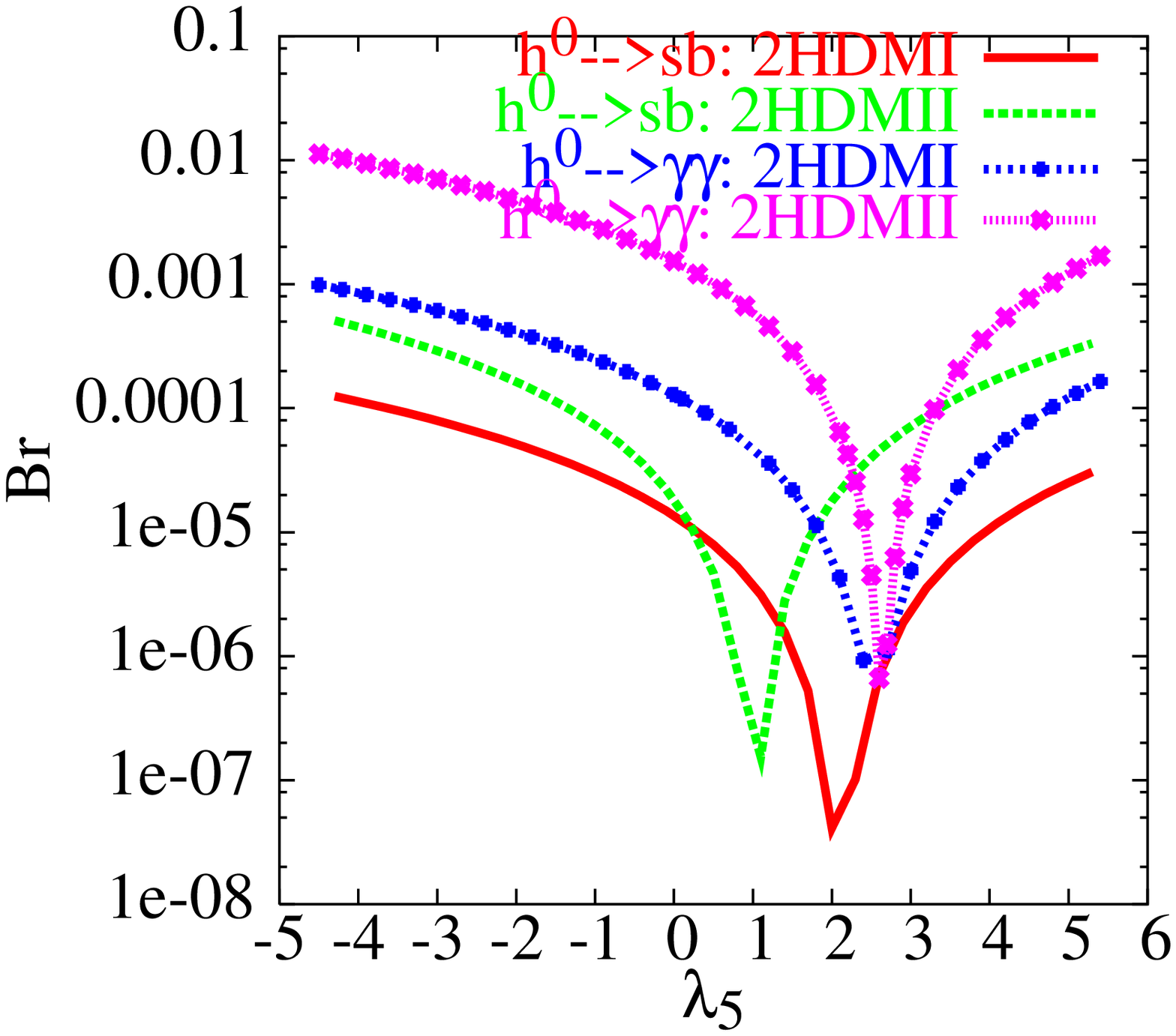}}  \hskip0.4cm
\epsfxsize2.8 in 
\epsffile{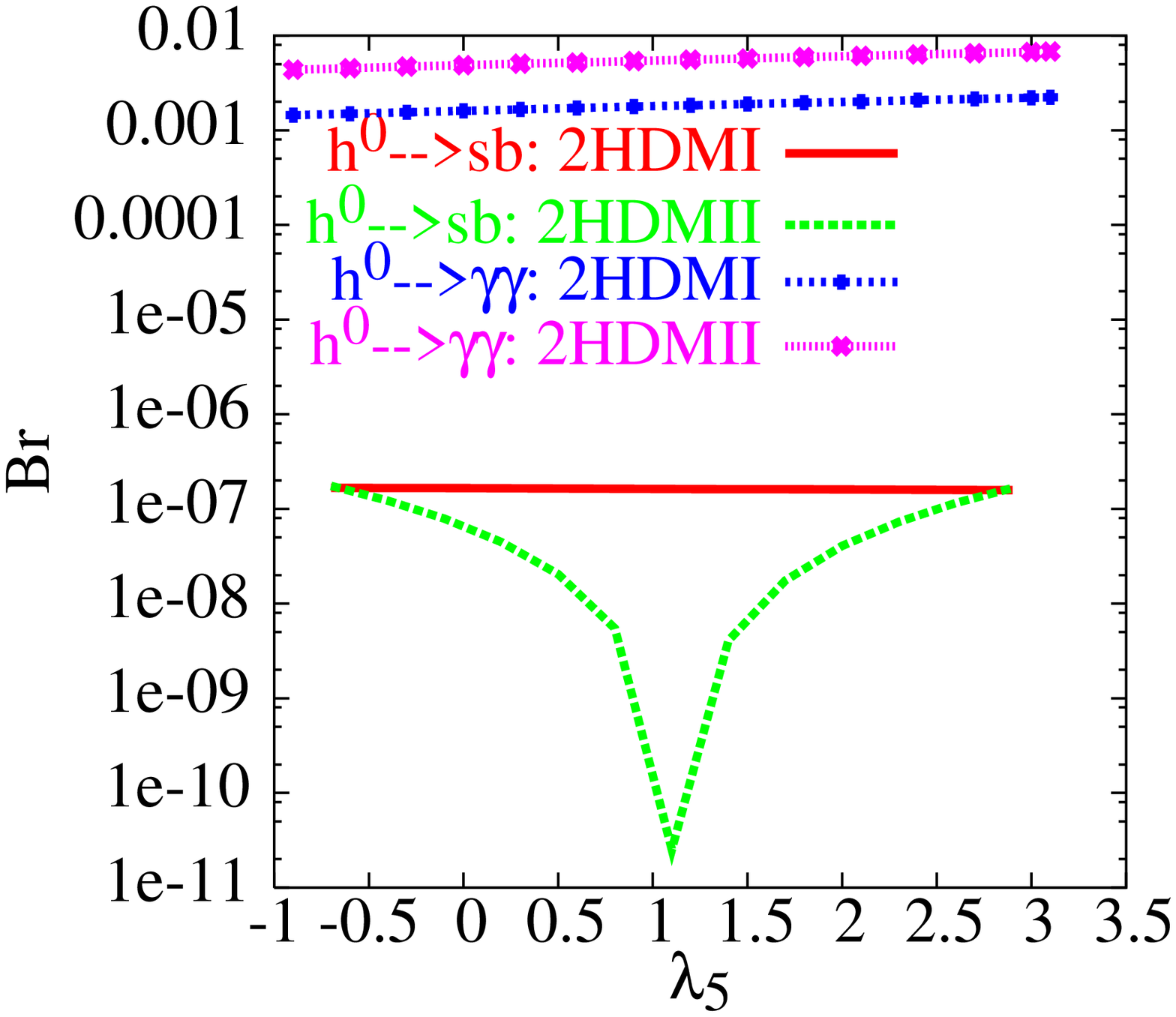} }
\smallskip\smallskip
\caption{$2\times Br(h^0\to \gamma \gamma)$ and $2\times Br(h^0 \to \bar{s}b)$
in 2HDM-I and II, $m_{H\pm}=100$ GeV, $\tan\beta=0.3$ (left) 
and $\tan\beta=5$ (right). All the other parameters are 
the same as in Fig.~(\ref{fig9}).}
\label{fig12}
\end{figure}

Large branching ratios
can be obtained for light charged Higgs mass. This can be seen 
in the left panel black and blue areas of Fig.~(\ref{fig9}) which correspond 
to small $\tan\beta=0.3$ and large $|\lambda_5|$. 
In those areas the coupling $h^0H^+H^-$ gets
its largest value (see also Fig.~(\ref{fig12})). 
In this case one can obtain branching ratio 
in the range: $10^{-4}< Br(h^0\to \bar{s}b) < 6\times
10^{-4}$ for $M_{H\pm}< 200 $ GeV , $\lambda_5\la -1.2$ and
$\lambda_5\ga 3$. For charged Higgs mass greater than 200 GeV,
there is also a region where the branching ratio can be in the range
$10^{-5} \to 10^{-4}$. This can be achieved by taking large and  
negative $\lambda_5\la -1$. In the case of positive $\lambda_5$
and $M_{H\pm}\ga 250$ GeV, the branching ratio decreases to a value
 $\la 10^{-5}$.

When $\tan\beta=1.5$, the coupling $h^0t\bar{t}$ is reduced,
and we are left only with a small region where the branching ratio 
$Br(h^0\to  \bar{s}b)$
is of the order $10^{-5}\to 10^{-4}$ for $M_{H\pm}\la 250 $ GeV
and large $|\lambda_5|\ga 5$. In both plots (left and right),
the coupling $h^0H^+H^-$ reaches its minimal value
in the region where $\lambda_5\approx 0\to 2$, which explains why 
the branching ratio is so small in this region. 

Now we turn to the case where $M_{H\pm}\neq M_A$, $\delta\rho\neq 0$.
We have performed a systematic scan over the full 2HDM parameters space
taking into account $\delta\rho$  and perturbativity constraints.
The maximum branching ratios found for $h^0\to \bar{s}b$ in 
2HDM-I and II are displayed in table~1. 
We show not only width and Br of
$h^0\to \bar{s}b$ but also the width and Br of 
$h^0\to \gamma \gamma$ for comparison. 
The total width of the Higgs
$\Gamma_{h^0}$ is also given. When  $\Gamma_{h^0}$
becomes comparable to the width of $h^0\to \bar{s}b$ and/or 
$h^0\to \gamma \gamma$, those decays widths have to be included
in the total width $\Gamma_{h^0}$  in order to
compute the Br$_{\bar{s}b}$ and Br$_{\gamma\gamma}$.

\begin{table}
\begin{center}
\begin{tabular}{||c|c|c|c|c|c|c|c|c|c|c|}
\hline
\hline
$ \begin{array}{c}
M_h \\
M_H
\end{array}\begin{array}{c}
M_{H\pm}\\
M_A
\end{array}$ & $ \begin{array}{c}
\sin\alpha \\
\tan\beta
\end{array}$&
$\lambda_5$ & $2\times$ Br$_{\bar{s}b}$ & $2\times$ $\Gamma_{\bar{s}b}$ 
& $2\times$ Br$_{\gamma\gamma}$
& $2\times \Gamma_{\gamma\gamma}$  & $\Gamma_h$
\\
\hline 
$ \begin{array}{c}
95 \\
340
\end{array} \begin{array}{c}
100 \\
110
\end{array}$  &
$ \begin{array}{c}
-.98 \\
.2
\end{array}$ &     6
& $ \begin{array}{c}
 10^{-3} \\
7\times 10^{-4} 
\end{array}$   &
$ \begin{array}{c}
6\times 10^{-6} \\
3\times 10^{-6}
\end{array}$ &
$ \begin{array}{c}
6\times 10^{-3} \\
6\times 10^{-3} 
\end{array}$ & 
$ \begin{array}{c}
3\times 10^{-5} \\
3\times 10^{-5} 
\end{array}$ &
$ \begin{array}{c}
5\times 10^{-3} \\
5\times 10^{-3} 
\end{array}$
\\
\hline 
$ \begin{array}{c}
140 \\
340 
\end{array}\begin{array}{c}
110 \\
100 
\end{array}$ &$ \begin{array}{c}
-.96 \\
.25
\end{array}$ &        6 &
$ \begin{array}{c}
4\times 10^{-4} \\
3\times 10^{-4}
\end{array}$ &
$ \begin{array}{c}
4\times 10^{-6} \\
2\times 10^{-6}
\end{array}$ &
$ \begin{array}{c}
10^{-2} \\
2\times 10^{-2}
\end{array}$ &
$ \begin{array}{c}
 10^{-4} \\
 10^{-4}
\end{array}$ &
$ \begin{array}{c}
9\times 10^{-3} \\
7\times 10^{-3}
\end{array}$ 
\\
\hline 
$ \begin{array}{c}
135 \\
160
\end{array}\begin{array}{c}
105 \\
240
\end{array}$ &$ \begin{array}{c}
-.98 \\
.46
\end{array}$&        -12 &
$ \begin{array}{c}
10^{-3} \\
2\times 10^{-4}
\end{array}$ &
$ \begin{array}{c}
2\times 10^{-6} \\
 10^{-6}
\end{array}$ &
$ \begin{array}{c}
7\times 10^{-3} \\
2\times 10^{-3}
\end{array}$ &
$ \begin{array}{c}
10^{-5} \\
10^{-5}
\end{array}$ &
$ \begin{array}{c}
2\times 10^{-3} \\
7\times 10^{-3}
\end{array}$ \\
\hline 
$ \begin{array}{c}
115 \\
250
\end{array}\begin{array}{c}
110 \\
190
\end{array}$ &$ \begin{array}{c}
.1 \\
.1
\end{array}$&        0 &
$ \begin{array}{c}
9\times 10^{-4} \\
 10^{-3} 
\end{array}$ &
$ \begin{array}{c}
5\times 10^{-4} \\
5\times 10^{-5} 
\end{array}$ &
$ \begin{array}{c}
2 \times10^{-4} \\
3 \times10^{-3} 
\end{array}$ &
$ \begin{array}{c}
10^{-4} \\
 10^{-4} 
\end{array}$ &
$ \begin{array}{c}
.55 \\
5\times 10^{-2} 
\end{array}$
\\
\hline 
$ \begin{array}{c}
110 \\
210
\end{array}\begin{array}{c}
105 \\
150
\end{array}$ &$ \begin{array}{c}
.18 \\
.1
\end{array}$&        0 &
$ \begin{array}{c}
9\times 10^{-4} \\
 10^{-3} 
\end{array}$ &
$ \begin{array}{c}
5\times 10^{-4} \\
5\times 10^{-5} 
\end{array}$ &
$ \begin{array}{c}
2 \times10^{-4} \\
3 \times10^{-3} 
\end{array}$ &
$ \begin{array}{c}
10^{-4} \\
10^{-4} 
\end{array}$ &
$ \begin{array}{c}
.52 \\
5\times 10^{-2} 
\end{array}$
\\
\hline
\end{tabular}
{\caption{Maximum Branching ratios of $h^0\to \bar{s}b$ 
in 2HDM-I and II and corresponding 2HDM parameters, all masses and
decay width are in GeV. In $Br$ and widths $\Gamma$ columns,
the upper row is for 2HDM-I and the down row is for 2HDM-II}}
\end{center}
\end{table}
\noindent
The first three columns of table~1 are for 2HDM parameters.
From 4th to 8th columns we give Br and widths. In those columns,
the upper row is for 2HDM-I and the down row is for 2HDM-II.\\
In 2HDM-I, Br$(h^0\to \bar{s}b)$ of the order $10^{-3}$
can be reached in the limit $\sin\alpha\to -0.98$ 
($\alpha\to -\pi/2$) and small
$\tan\beta \leq 0.5$. In fact, this limit ($\alpha\to -\pi/2$)
is very close to fermiophobic scenario $\alpha=\pm \pi/2$.
In the fermiophobic limit, all couplings of $h^0$ to down 
quarks and leptons are suppressed eq.~(\ref{coupl2}). In this limit,
$h^0\bar{t}t$ is also suppressed eq.~(\ref{coupl1}). 
The width of light Higgs $h^0$ ($M_h<160$ GeV) is then very tiny
in the limit $\sin\alpha\to -0.98$. This tiny width together with 
large $h^0H^+H^-$ are the sources of enhancement of  
the Br$(h^0\to \bar{s}b)$ to $10^{-3}$ level. 
This can be seen in the first, second and third lines of table~1\\
In 2HDM-II, the couplings of $h^0$ to down 
quarks and leptons are suppressed for $\sin\alpha\approx 0.1$ 
eq.~(\ref{coupl3}).
Hence, the width of light Higgs ($M_h<160$ GeV) 
is very tiny in the limit $\sin\alpha\approx 0.1$.
Moreover, in this limit, the coupling $h^0\bar{t}t$ is enhanced
in both models 2HDM-I and II. The decay width $\Gamma(h^0\to
{\bar{s}b})$ which was $\approx 10^{-6}$ for $\sin\alpha=-0.98$ 
is of the order $\approx 10^{-5}$ for $\sin\alpha=0.1$.
Consequently, the Br$(h^0\to \bar{s}b)$ reaches $10^{-3}$.\\
In this scenario, as one can see from table~1, the 
Br$(h^0\to \gamma\gamma)$ in 2HDM-I is $2\times 10^{-4}$ which is
smaller than Br$(h^0\to \bar{s}b)=9\times 10^{-4}$. This is mainly due
to the fact that the trilinear coupling $h^0H^+H^-$ is very suppressed
in this scenario (see more details in next section).\\
As one can see from the last line of the table~1, there exist also 
values of $\sin\alpha=0.18$, far from fermiophobic 
scenario but with small $\lambda_5=0$, where 
Br$(h^0\to \bar{s}b)$ can be of the order $10^{-3}$.\\

\subsection{Can $h^0\to \bar{s}b$ compete with $h^0\to \gamma \gamma$?}
It is well known that the decay $h^0\to \gamma \gamma$
is loop induced and so is suppressed. In the SM, the branching ratio
$Br(H^{SM}\to \gamma \gamma)$ is about $\approx 10^{-3}$ 
for Higgs mass in the range $M_H=100\to 160$ GeV.
Hence, with maximum branching ratio for $h^0\to \bar{s}b$ of the order 
$1\times 10^{-4}\to 6 \times 10^{-4}$
in 2HDM-I or II, it is legitimate to compare $h^0\to \gamma \gamma$ 
and $h^0\to \bar{s}b$ in 2HDM-I or II. Of course, even if  
$h^0\to \bar{s}b$ and $h^0\to \gamma \gamma$ has a competitive 
branching ratio, we should keep in mind that $h^0\to \gamma \gamma$
decay has a clear signature while the FCNC decay  $h^0\to \bar{s}b$
has not.

We illustrate in Fig.~(\ref{fig12}) the branching ratio for 
$h^0\to \bar{s}b$ and $h^0\to \gamma \gamma$ both in 2HDM-I and II.
The charged Higgs mass is fixed to 100 GeV. 
It is clear that in the case $\tan\beta=0.3$ $h^0\to \gamma \gamma$ 
is about one order of magnitude bigger than $h^0\to \bar{s}b$.
While, in the case of $\tan\beta=5$ $h^0\to \gamma \gamma$ 
is more than four orders of magnitude bigger than $h^0\to \bar{s}b$.
This is because at $\tan\beta=0.3$ (resp  $\tan\beta=5$) 
the W loop are suppressed by a
factor $h^0W^+W^-\propto \sin(\beta-\alpha)\approx 0.2$
(resp enhanced by $h^0W^+W^-\propto \sin(\beta-\alpha)\approx  0.96$).
All the dips observed in the plots correspond to the minimum of the 
coupling $h^0H^-H^+$. Those dips are not located at the same
$\lambda_5$, this is due to a destructive interference 
with others diagrams. When $h^0H^-H^+$ coupling is very suppressed,
it may be possible that the Br$(h^0\to \bar{s}b)$ could be higher
than Br$(h^0\to \gamma \gamma)$ as it can be seen
both in the left plot of Fig.~(\ref{fig12}) for $\lambda_5=2.5$ and 
 in table~1 for  $\sin\alpha=0.1$ in 2HDM-II.\\
However, even if Br$(h^0\to \gamma \gamma)$ and Br$(h^0\to \bar{s}b)$
become comparable,
we should keep in mind that $h^0\to \gamma \gamma$ 
has a very clear signature while $h^0\to \bar{s}b$ does not.

An interesting feature of the 2HDM-I, is its fermiophobic scenario.
The light CP-even Higgs $h^0$ of the 2HDM-I 
is fermiophobic in the limit $\alpha\to \pi/2$, 
all $h^0$ couplings to fermions vanishes for $\alpha=\pi/2$ \cite{andrew,Gun}.
If $h^0$, with a mass in the range $100\to 160$ GeV, is fermiophobic
the dominant decay mode is $h^0\to \gamma \gamma$. 
It has been shown in Ref.~\cite{santos} that in the fermiophobic
limit, the branching ratio of the one loop induced decay 
$h^0\to \bar{b}b$\footnote{In fact, in the 2HDM, not only the
 coupling $h^0 \gamma \gamma$ and 
$h^0 \gamma Z$ \cite{ITPM} can have non decoupling effects, 
but also one loop contribution to $h^0\bar{s}b$ \cite{AHPC} and 
$h^0h^0h^0$ \cite{okada}.} is  
below $10\% \to 30\%$.
As the decay $h^0\to \bar{s}b$ is concerned, we have checked by
systematic scan that in the fermiophobic limit, the decay width 
of $h^0\to \gamma \gamma$ is more than one order of magnitude bigger than 
the width of $h^0\to \bar{s}b$.

\subsection{$H^0 \to {\bar{s}}b$}
\begin{figure}[t!]
\smallskip\smallskip 
\vskip-.1cm
\centerline{{
\epsfxsize2.8 in 
\epsffile{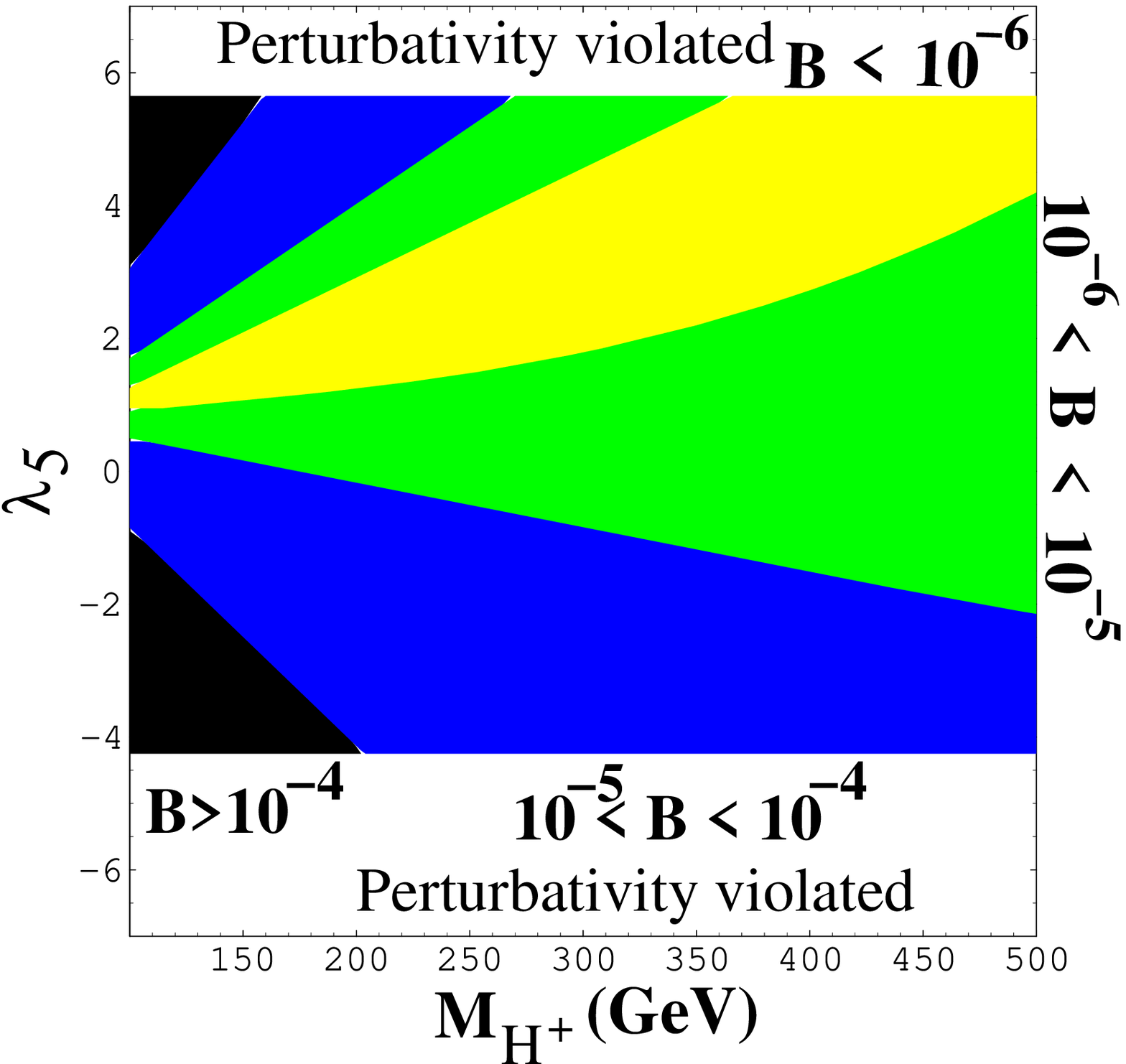 }}  \hskip0.4cm
\epsfxsize2.8 in 
\epsffile{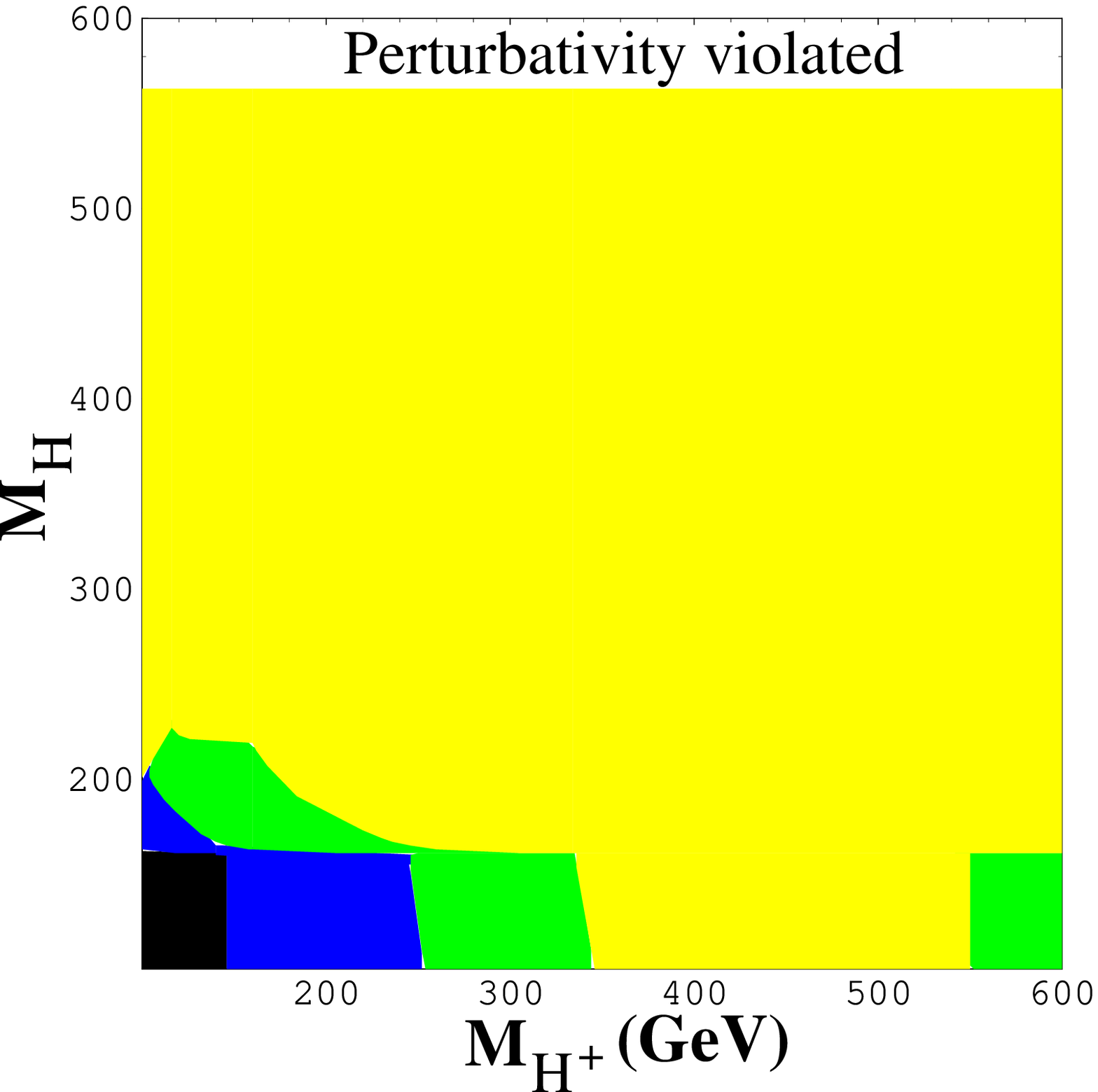 }}
\smallskip\smallskip
\caption{Contours for $2\times Br(H^0\to  \bar{s}b)$ in 
2HDM-II in the plan ($M_{H\pm}$, $\lambda_5$) $M_H=140$ GeV (left),
 ($M_{H\pm}$, $M_H$) $\lambda_5=5$ (right)  with
$\tan\beta=0.3$, $M_h=110$ GeV, $\sin\alpha=0.95$ and $M_{A^0}=M_{H\pm}$}
\label{fig10}
\end{figure}
We now  discuss the heavy CP-even decay $H^0\to \bar{s}b$. 
Our numerical results are shown in Fig.~(\ref{fig10}). To maximize
the coupling $H^0\bar{t}t$, we choose of course small $\tan\beta\approx
0.3$ and large $\sin\alpha\approx 0.95$. In the right plot of 
Fig.~(\ref{fig10}), we show contour plots for 
$Br(H^0\to \bar{s}b)$ in the plane $(M_{H\pm},\lambda_5)$
for $M_{H}=140$ GeV. For CP-even Higgs mass 140 GeV, $H^0\to W^+W^- $,
$H^0\to ZZ $, $H^0\to \bar{t}t$, 
$H^0\to A^0 Z$ and $H^0\to H_i H_j$  are not yet
open, and so the width is narrow. In particular, for the set of 
parameters fixed here:
$M_h=110$ GeV, $\sin\alpha=0.95$ and $M_{A^0}=M_{H\pm}$, 
the width is $73\times 10^{-4}$ GeV. \\
The behavior is similar to what we obtain
for $Br(h^0\to \bar{s}b)$. 
In the black regions (large $\lambda_5$), the coupling
$H^0H^+H^-$ is maximal while for $\lambda_5\in [0,2]$
$H^0H^+H^-$ is minimal. 
In the black region the branching ratio of 
$Br(H^0\to \bar{s}b)$ can reach $\approx 7\times 10^{-4}$.
From the left panel of Fig.~(\ref{fig10}), it is evident that 
there is a relatively large region in the plane ($M_{H\pm},\lambda_5$) 
where the $Br(H^0\to \bar{s}b) \ga 10^{-5}$.

\begin{table}[t!]
\begin{center}
\begin{tabular}{||c|c|c|c|c|c|c|c|c|c|c|}
\hline
\hline
$ \begin{array}{c}
M_h \\
M_H
\end{array}\begin{array}{c}
M_{H\pm}\\
M_A
\end{array}$ & $ \begin{array}{c}
\sin\alpha \\
\tan\beta
\end{array}$ &
$\lambda_5$ & $2\times$ Br$_{\bar{s}b}$ & $2\times$ $\Gamma_{\bar{s}b}$ 
& $2\times$ Br$_{\gamma\gamma}$
& $2\times \Gamma_{\gamma\gamma}$  & $\Gamma_H$
\\
\hline 
$ \begin{array}{c}
115 \\
155 
\end{array}\begin{array}{c}
142 \\
183 
\end{array}$    &  $ \begin{array}{c}
.1 \\
.5 
\end{array}$ &
        -12 & $ \begin{array}{c}
10^{-3} \\
4\times 10^{-5}
\end{array}$ 
&  $ \begin{array}{c}
5\times 10^{-7} \\
4\times 10^{-7}
\end{array}$ &
$ \begin{array}{c}
3\times 10^{-2} \\
10^{-3}
\end{array}$ & 
$ \begin{array}{c}
 10^{-5} \\
 10^{-5}
\end{array}$&
$ \begin{array}{c}
4\times 10^{-4} \\
8\times 10^{-3}
\end{array}$
\\
\hline 
$ \begin{array}{c}
100 \\
155
\end{array}\begin{array}{c}
115 \\
103
\end{array}$ & $ \begin{array}{c}
.1 \\
.34
\end{array} $&
    -6 & $ \begin{array}{c}
2\times 10^{-3} \\
8\times 10^{-5}
\end{array}$     &  $ \begin{array}{c}
10^{-6} \\
6\times 10^{-7}
\end{array}$ & 
$ \begin{array}{c}
2\times 10^{-2} \\
2\times 10^{-3}
\end{array}$ & $ \begin{array}{c}
2\times 10^{-5} \\
2\times 10^{-5}
\end{array}$ & $ \begin{array}{c}
7\times 10^{-4} \\
8\times 10^{-3}
\end{array}$
\\
\hline
$ \begin{array}{c}
100 \\
155
\end{array} \begin{array}{c}
120 \\
140
\end{array}$  & $ \begin{array}{c}
.08 \\
.45
\end{array}$ &
    12 &  $ \begin{array}{c}
2\times 10^{-3} \\
8\times 10^{-5}
\end{array}$     &  $ \begin{array}{c}
7\times 10^{-7} \\
6\times 10^{-7}
\end{array}$ & 
$ \begin{array}{c}
.73 \\
3\times 10^{-2}
\end{array}$ & $ \begin{array}{c}
2\times 10^{-4} \\
2\times 10^{-4}
\end{array}$ & $ \begin{array}{c}
3\times 10^{-4} \\
8\times 10^{-3}
\end{array}$    
\\
\hline
$ \begin{array}{c}
100 \\
125
\end{array} \begin{array}{c}
115 \\
103
\end{array}$  & 
$ \begin{array}{c}
-.98 \\
.1
\end{array} $ & 0  &   
$ \begin{array}{c}
9\times 10^{-4} \\
 10^{-3}
\end{array}$  & 
$ \begin{array}{c}
5\times 10^{-4} \\
6\times 10^{-5}
\end{array}$  &
$ \begin{array}{c}
3\times 10^{-4} \\
4\times 10^{-3}
\end{array}$  & 
$ \begin{array}{c}
2\times 10^{-4} \\
2\times 10^{-4}
\end{array}$ &$ \begin{array}{c}
.58 \\
5\times 10^{-2}
\end{array}$
\\
\hline
$ \begin{array}{c}
100 \\
130
\end{array} \begin{array}{c}
110 \\
300
\end{array}$  &$ \begin{array}{c} .9 \\ .1 \end{array}$  &  0 &   
$ \begin{array}{c}
10^{-3} \\  10^{-3}
\end{array}$  & $ \begin{array}{c}
5\times 10^{-4} \\ 6\times 10^{-5}
\end{array}$ 
&$ \begin{array}{c}
2\times 10^{-4} \\ 2\times 10^{-3}
\end{array}$ &
$ \begin{array}{c}
 10^{-4} \\ 10^{-4}
\end{array}$ & $ \begin{array}{c}
 .51\\ 5\times 10^{-2} 
\end{array}$ 
\\
\hline
$ \begin{array}{c}
100 \\
145
\end{array} \begin{array}{c}
115 \\
103
\end{array}$  &$ \begin{array}{c} -.58 \\ .1 \end{array}$  &  0 &   
$ \begin{array}{c}
9\times 10^{-4} \\  10^{-3}
\end{array}$  & $ \begin{array}{c}
2\times 10^{-4} \\ 3\times 10^{-5}
\end{array}$ 
&$ \begin{array}{c}
10^{-3} \\ 10^{-2}
\end{array}$ &
$ \begin{array}{c}
2\times 10^{-4} \\ 3 \times 10^{-4}
\end{array}$ & $ \begin{array}{c}
 .24\\ 3\times 10^{-2} 
\end{array}$
\\ \hline
\end{tabular}
{\caption{Maximum Branching ratios of $H^0\to \bar{s}b$ 
in 2HDM-I and II and corresponding 2HDM parameters, all masses and
decay width are in GeV. In $Br$ and widths $\Gamma$ columns,
the upper row is for 2HDM-I and the down row is for 2HDM-II}}
\end{center}
\end{table}
In the right panel of Fig.~(\ref{fig10}), we show $Br(H^0\to
\bar{s}b)$ in the plan $(M_H,M_{H\pm})$ for $\lambda_5=5$.
One can see that when CP-even mass $M_H < 2 M_W$, the decay $H^0\to
W^+W^-$ is not yet open. The width $\Gamma_{H^0}$ is narrow, 
and so the branching ratio is large. For $M_{H\pm}<250$ GeV 
and $M_H< 2 M_W$, one can have $Br(H^0\to
\bar{s}{b})\ga 10^{-5}$. Once the CP-even Higgs mass 
$M_H > 2 M_W$, the decay $H^0\to W^+W^-$ is open, and
the width is larger than $5\times 10^{-2}$ GeV. The Branching ratio
$Br(H^0\to \bar{s}{b})$ is then reduced. As it can be seen
from the right plot, for $M_H \ga 220$ GeV, the branching ratio $Br(H^0\to
\bar{s}{b})$ is less than $\la 10^{-6}$.

In case of 2HDM-I, both $h^0t\bar{t}$, $H^0\bar{t}t$  
and $(H^-\bar{b}t)_R$ couplings are the same as in 2HDM-II, 
while $(H^-\bar{b}t)_L$ 
which is proportional to $M_b\tan\beta$ in 2HDM-II is now 
proportional to $M_b/\tan\beta$. For small $\tan\beta\approx 0.3$,
both $Br(h^0\to \bar{s}b)$ and 
$Br(H^0\to \bar{s}b)$ are of  the same order as in
2HDM-II, while for large $\tan\beta$ those Branching ratios are less
than about $\approx 10^{-6}$.

In case where $M_{H\pm}\neq M_A$ ($\delta\rho\neq 0$),
we present our results of maximum branching ratios of $H^0\to \bar{s}b$
in the table~2. It turns out that in 2HDM-I (resp 2HDM-II), 
Br$(H^0\to \bar{s}b)$ reach $10^{-3}$ for small
$\sin\alpha\approx 0.1$  (resp large $|\sin\alpha|\approx 0.9$).
The interpretation is the same as in the case of light CP even Higgs
$h^0$. In 2HDM-I (resp 2HDM-II), the couplings of $H^0$ to down 
quarks and leptons are suppressed for $\sin\alpha\approx 0.1$
(resp $|\sin\alpha|\approx 0.9$). In those cases the total Higgs width
is very tiny and so the branching ratio of $H^0\to \bar{s}b$ is 
enhanced.\\
Of course, Br$(H^0\to \bar{s}b)$ reach $10^{-3}$ only for light
charged Higgs, which is strongly disfavored by 
$b\to s\gamma$ constraint \cite{bsg} in 2HDM-II.\\
From table~2, one can see also that in 2HDM-II and for 
$\sin\alpha\approx \{0.9,-0.98\}$ the Br$(H^0\to \bar{s}b)$
and Br$(H^0\to \gamma \gamma)$ are of comparable size.
This is again mainly due to the suppression of the 
coupling $H^0H^+H^-$  in those limits.\\
As in the case of light CP-even Higgs $h^0$, there exist 
values of $\sin\alpha=-0.58$ far from 
fermiophobic limit with small $\lambda_5=0$ where 
Br$(H^0\to \bar{s}b)$ can reach $10^{-3}$.

\subsection{$A^0 \to {\bar{s}}b$}
Let us now look at 2HDM contribution to $A^0\to \bar{s}b$. Since $A^0$ 
is CP-odd, it does not couple to a pair of charged Higgs. The 
only pure trilinear scalar coupling which contributes to 
$A^0\to \bar{s}b$ is $A^0H^\pm G^\mp$ eq.~(\ref{scalar5}).
Unlike the couplings $H^0H^+H^-$ and $h^0H^+H^-$ 
eqs~(\ref{scalar1},\ref{scalar3}), which depend both on Higgs masses,
$\tan\beta$ as well as the soft breaking term $\lambda_5$,
the coupling $A^0H^\pm G^\mp$ depends only on the splitting 
$M_{H\pm}^2-M_A^2$. As mentioned above, such splitting 
should not be too large, otherwise the $\delta\rho$ constraint is 
not satisfied. As one can read from eqs.~(\ref{coupl1},\ref{coupl5}),
the couplings $A^0\bar{t}t$ and  $(H^-\bar{b}t)_R$
are proportional to $M_t/\tan\beta$. Hence enhancement is expected at
small $\tan\beta$.

\begin{figure}[t!]
\smallskip\smallskip 
\vskip-3.8cm
\centerline{{
\epsfxsize2.8 in 
\epsffile{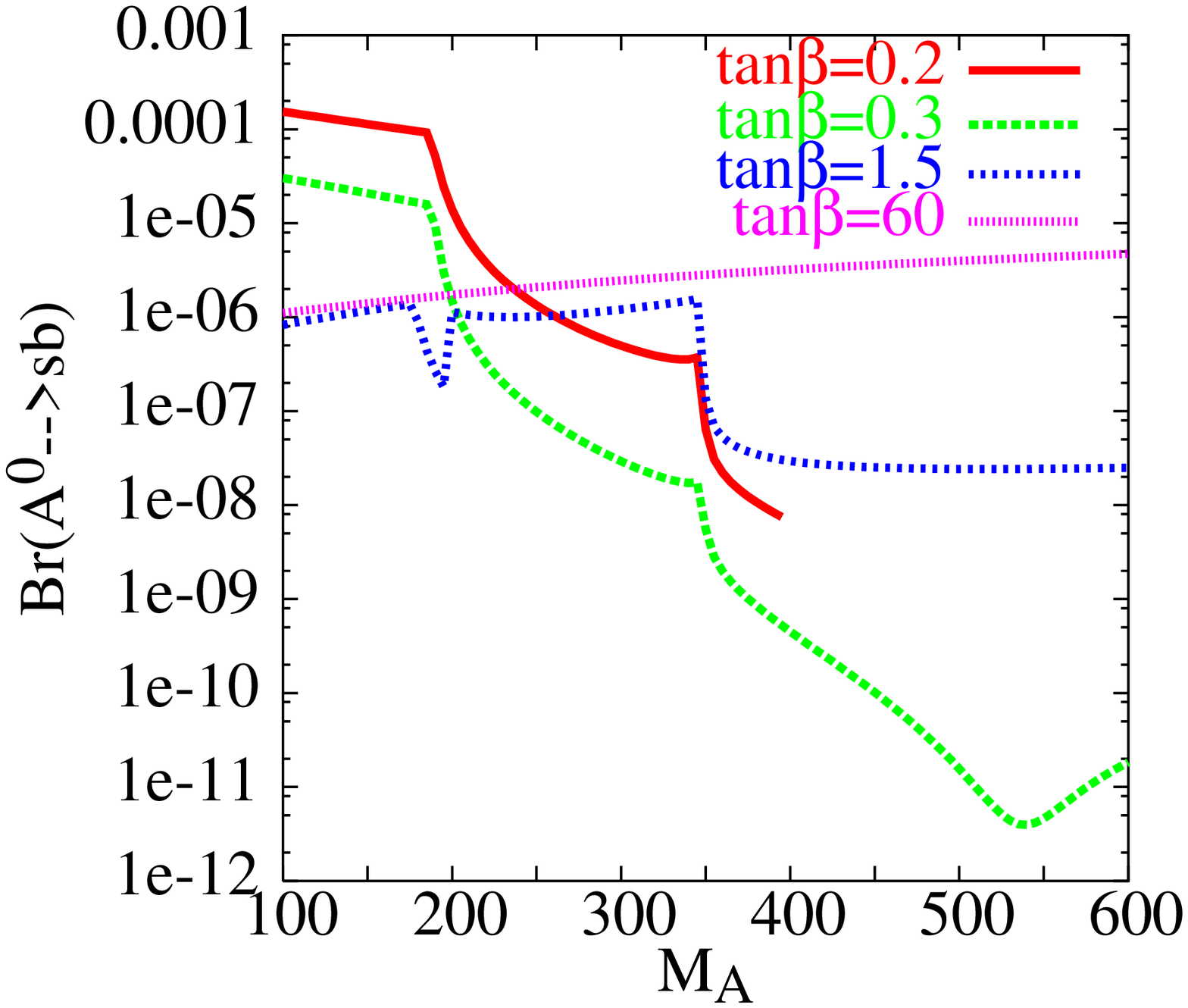}}  \hskip0.4cm
\epsfxsize2.8 in 
\epsffile{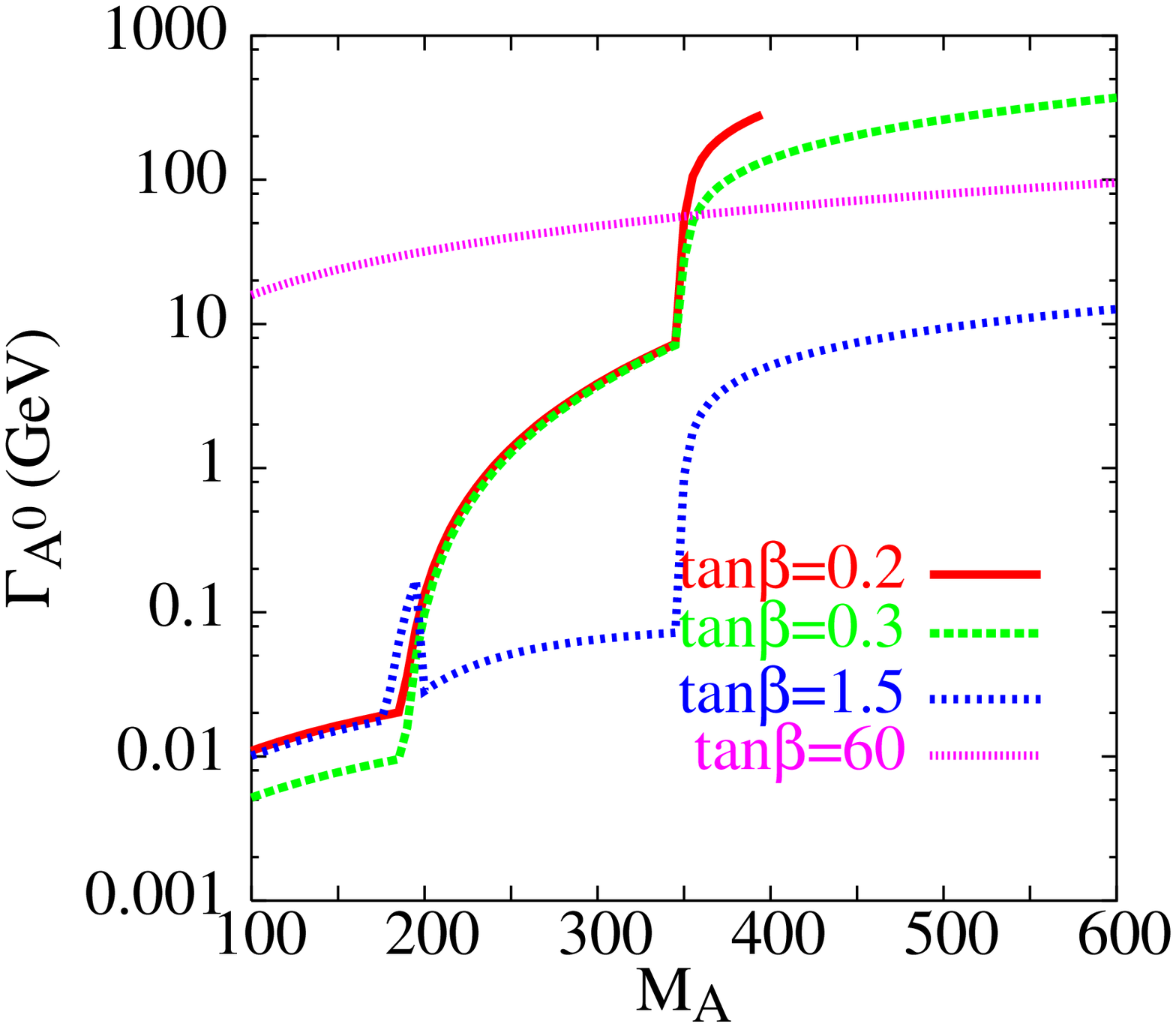} }
\smallskip\smallskip
\caption{$2\times Br(A^0\to \bar{s}b)$ (left) and CP-odd $A^0$ width
  $\Gamma_{A^0}$ (right) as function of $M_A$ 
for several values of $\tan\beta$}
\label{fig11}
\end{figure}
As we stressed before, our 2HDM parameters in this case are:
$\tan\beta$, $M_A$ and $M_{H\pm}$.
For simplification, we use the MSSM sum-rules to fix
charged Higgs mass and $\alpha$ by using 
$\tan\beta$, CP-odd mass $M_A$ and 
a SUSY scale which we take at 1 TeV. 
CP-odd mass will be varied from 100 GeV to 600 GeV
without worrying about perturbativity. 
$\tan\beta$ is taken to be $\ga 0.1$.\\
We present our numerical results for 
$A^0\to \bar{s}b$ in 2HDM-II in Fig.~(\ref{fig11}).
As can be seen from the left plot, the Branching ratio $Br(A^0\to
\bar{s}b)$ is greater than $10^{-5}$ only for small $\tan\beta\approx
0.1\to 0.35$ and light $M_A$ and $M_{H\pm}$. For light $M_A\la 200$
GeV and low $\tan\beta \la 1$, the width of $A^0$ is still small 
and so the branching ratio
is enhanced. For $M_A\ga 200$ GeV, the decay $A^0\to h^0 Z$ is open
and the decay width $\Gamma_{A^0}$ increases. Therefore, the branching
ratio is reduced.
Note that for $\tan\beta=0.2$, we cut off the curve at $M_A\approx 400$
GeV where the width $\Gamma_{A^0}$ starts to be greater than $M_A$.
At large $\tan\beta$, due to the bottom Yukawa coupling, 
 both the partial width $\Gamma(A^0\to \bar{s}b)$ and total
width $\Gamma_{A^0}$ are enhanced, and the branching ratio is saturated
in the range $[10^{-6},10^{-5}]$.\\
The situation is almost the same in 2HDM-I.

\section{Conclusions}
In the framework of the 2HDM with natural flavor conservation,
we have studied various Higgs FCNC $\Phi\to \bar{s}b$. 
The study has been carried out
taking into account the experimental constraint on the $\rho$ parameter 
and also perturbativity constraints on all the scalar 
quartic couplings $\lambda_i$.
Numerical results for the branching ratios have been discussed.
We emphasized the effect coming from both top and bottom Yukawa couplings
and pure trilinear scalar couplings such as $h^0H^+H^-$ and $H^0H^+H^-$.
\\
\noindent
We have shown that, in 2HDM-I and 2HDM-II,  
the branching ratios of Higgs FCNC 
$\{h^0 ,H^0, A^0\}\to \bar{s}b$ are enhanced to the range of
$10^{-4} \to 7\times 10^{-4} $
for small $\tan\beta$, rather light charged Higgs boson and
large soft breaking term $\lambda_5$.
The branching ratio of $Br(\{h^0 ,H^0\}\to \bar{s}b)$ can be pushed to 
$10^{-3}$ level when $\sin\alpha$ is close to fermiophobic
limit ($\sin\alpha\approx -0.98$)
or  $\sin\alpha\approx 0.1$ and even for $\sin\alpha$ far from those
limits but with small $\lambda_5=0$.\\
Charged Higgs mass of 2HDM-I is
not constrained by $b\to s \gamma$,  
$Br(\{h^0 ,H^0\}\to \bar{s}b)$ can be of 
the order $10^{-4}\to10^{-3}$  for light charged Higgs which is 
comparable to size of SUSY predictions \cite{maria1,bdgs}. 
Those branching ratios rates, could still leads to large 
number of events at LHC \cite{2hdm4}.\\
In 2HDM-II with $b\to s \gamma$ constraint, branching ratios of 
$\{h^0 ,H^0\}\to \bar{s}b$ are smaller than $10^{-5}$
(resp $10^{-4}$) for $\tan\beta > 1$ (resp $\tan\beta < 1$). \\
In the case of light CP-even $m_{h^0}\approx 100\to  160$ GeV,
we have also shown that the branching ratio of $Br(h^0\to \bar{s}b)$
is well below $Br(h^0\to \gamma \gamma)$ in most of the case. 
This is also the case in the fermiophobic scenario of 2HDM-I.\\
One interesting scenario is that both $Br(h^0\to \gamma \gamma)$
and $Br(h^0\to \bar{s}b)$ 
develop a dips for some $\lambda_5$ (see Fig.~4). Those dips are 
not located at the same $\lambda_5$ due to the presence 
of diagrams which contribute
to $h^0\to \bar{s}b$ but not to $h^0\to \gamma \gamma$.
The dip for $Br(h^0\to \bar{s}b)$ is located for $\lambda_5=1$
while for $Br(h^0\to \gamma \gamma)$ it is located for
$\lambda_5\approx 2.5$. For $\lambda_5\approx 2.5$, we are already
away from $Br(h^0\to \bar{s}b)$ dip, the 
$Br(h^0\to \bar{s}b)$ is slightly higher than $Br(h^0\to \gamma \gamma)$.

\vspace{1cm}
\noindent
{\Large \bf Acknowledgments}
This work was done within the framework of the 
Associate Scheme of ICTP. Thanks to Thomas Hahn for his help.
We also want to thank Andrew Akeroyd for discussions and for reading the 
manuscript.

\end{document}

%% file: hsbb.tex
\unitlength=1bp%

\begin{feynartspicture}(432,504)(6,6.3)

\FADiagram{d$_1$}
\FAProp(0.,10.)(6.5,10.)(0.,){/ScalarDash}{0}
\FALabel(3.25,9.18)[t]{$\Phi$}
\FAProp(20.,15.)(13.,14.)(0.,){/Straight}{-1}
\FALabel(16.2808,15.5544)[b]{$f_1$}
\FAProp(20.,5.)(13.,6.)(0.,){/Straight}{1}
\FALabel(16.2808,4.44558)[t]{$f_2$}
\FAProp(6.5,10.)(13.,14.)(0.,){/Straight}{1}
\FALabel(9.20801,13.1807)[br]{$f^\prime_i$}
\FAProp(6.5,10.)(13.,6.)(0.,){/Straight}{-1}
\FALabel(9.20801,6.81927)[tr]{$f^\prime_i$}
\FAProp(13.,14.)(13.,6.)(0.,){/ScalarDash}{1}
\FALabel(14.274,10.)[l]{$G$}
\FAVert(6.5,10.){0}
\FAVert(13.,14.){0}
\FAVert(13.,6.){0}

\FADiagram{d$_2$}
\FAProp(0.,10.)(6.5,10.)(0.,){/ScalarDash}{0}
\FALabel(3.25,9.18)[t]{$\Phi$}
\FAProp(20.,15.)(13.,14.)(0.,){/Straight}{-1}
\FALabel(16.2808,15.5544)[b]{$f_1$}
\FAProp(20.,5.)(13.,6.)(0.,){/Straight}{1}
\FALabel(16.2808,4.44558)[t]{$f_2$}
\FAProp(6.5,10.)(13.,14.)(0.,){/ScalarDash}{-1}
\FALabel(9.20801,13.1807)[br]{$G$}
\FAProp(6.5,10.)(13.,6.)(0.,){/ScalarDash}{1}
\FALabel(9.20801,6.81927)[tr]{$G$}
\FAProp(13.,14.)(13.,6.)(0.,){/Straight}{-1}
\FALabel(14.274,10.)[l]{$f^\prime_i$}
\FAVert(6.5,10.){0}
\FAVert(13.,14.){0}
\FAVert(13.,6.){0}

\FADiagram{d$_3$}
\FAProp(0.,10.)(6.5,10.)(0.,){/ScalarDash}{0}
\FALabel(3.25,9.18)[t]{$\Phi$}
\FAProp(20.,15.)(13.,14.)(0.,){/Straight}{-1}
\FALabel(16.2808,15.5544)[b]{$f_1$}
\FAProp(20.,5.)(13.,6.)(0.,){/Straight}{1}
\FALabel(16.2808,4.44558)[t]{$f_2$}
\FAProp(6.5,10.)(13.,14.)(0.,){/Straight}{1}
\FALabel(9.20801,13.1807)[br]{$f^\prime_i$}
\FAProp(6.5,10.)(13.,6.)(0.,){/Straight}{-1}
\FALabel(9.20801,6.81927)[tr]{$f^\prime_i$}
\FAProp(13.,14.)(13.,6.)(0.,){/Sine}{1}
\FALabel(14.274,10.)[l]{$W$}
\FAVert(6.5,10.){0}
\FAVert(13.,14.){0}
\FAVert(13.,6.){0}

\FADiagram{d$_4$}
\FAProp(0.,10.)(6.5,10.)(0.,){/ScalarDash}{0}
\FALabel(3.25,9.18)[t]{$\Phi$}
\FAProp(20.,15.)(13.,14.)(0.,){/Straight}{-1}
\FALabel(16.2808,15.5544)[b]{$f_1$}
\FAProp(20.,5.)(13.,6.)(0.,){/Straight}{1}
\FALabel(16.2808,4.44558)[t]{$f_2$}
\FAProp(6.5,10.)(13.,14.)(0.,){/ScalarDash}{-1}
\FALabel(9.20801,13.1807)[br]{$G$}
\FAProp(6.5,10.)(13.,6.)(0.,){/Sine}{1}
\FALabel(9.20801,6.81927)[tr]{$W$}
\FAProp(13.,14.)(13.,6.)(0.,){/Straight}{-1}
\FALabel(14.274,10.)[l]{$f^\prime_i$}
\FAVert(6.5,10.){0}
\FAVert(13.,14.){0}
\FAVert(13.,6.){0}

\FADiagram{d$_5$}
\FAProp(0.,10.)(6.5,10.)(0.,){/ScalarDash}{0}
\FALabel(3.25,9.18)[t]{$\Phi$}
\FAProp(20.,15.)(13.,14.)(0.,){/Straight}{-1}
\FALabel(16.2808,15.5544)[b]{$f_1$}
\FAProp(20.,5.)(13.,6.)(0.,){/Straight}{1}
\FALabel(16.2808,4.44558)[t]{$f_2$}
\FAProp(6.5,10.)(13.,14.)(0.,){/Sine}{-1}
\FALabel(9.20801,13.1807)[br]{$W$}
\FAProp(6.5,10.)(13.,6.)(0.,){/ScalarDash}{1}
\FALabel(9.20801,6.81927)[tr]{$G$}
\FAProp(13.,14.)(13.,6.)(0.,){/Straight}{-1}
\FALabel(14.274,10.)[l]{$f^\prime_i$}
\FAVert(6.5,10.){0}
\FAVert(13.,14.){0}
\FAVert(13.,6.){0}

\FADiagram{d$_6$}
\FAProp(0.,10.)(6.5,10.)(0.,){/ScalarDash}{0}
\FALabel(3.25,9.18)[t]{$\Phi$}
\FAProp(20.,15.)(13.,14.)(0.,){/Straight}{-1}
\FALabel(16.2808,15.5544)[b]{$f_1$}
\FAProp(20.,5.)(13.,6.)(0.,){/Straight}{1}
\FALabel(16.2808,4.44558)[t]{$f_2$}
\FAProp(6.5,10.)(13.,14.)(0.,){/Sine}{-1}
\FALabel(9.20801,13.1807)[br]{$W$}
\FAProp(6.5,10.)(13.,6.)(0.,){/Sine}{1}
\FALabel(9.20801,6.81927)[tr]{$W$}
\FAProp(13.,14.)(13.,6.)(0.,){/Straight}{-1}
\FALabel(14.274,10.)[l]{$f^\prime_i$}
\FAVert(6.5,10.){0}
\FAVert(13.,14.){0}
\FAVert(13.,6.){0}

\FADiagram{d$_7$}
\FAProp(0.,10.)(11.,10.)(0.,){/ScalarDash}{0}
\FALabel(5.5,9.18)[t]{$\Phi$}
\FAProp(20.,15.)(11.,10.)(0.,){/Straight}{-1}
\FALabel(15.2273,13.3749)[br]{$f_1$}
\FAProp(20.,5.)(17.3,6.5)(0.,){/Straight}{1}
\FALabel(18.9227,6.62494)[bl]{$f_2$}
\FAProp(11.,10.)(13.7,8.5)(0.,){/Straight}{-1}
\FALabel(12.0773,8.37506)[tr]{$f_1$}
\FAProp(17.3,6.5)(13.7,8.5)(-0.8,){/Straight}{1}
\FALabel(14.4273,5.18506)[tr]{$f^\prime_i$}
\FAProp(17.3,6.5)(13.7,8.5)(0.8,){/ScalarDash}{-1}
\FALabel(16.5727,9.81494)[bl]{$G$}
\FAVert(11.,10.){0}
\FAVert(17.3,6.5){0}
\FAVert(13.7,8.5){0}

\FADiagram{d$_{8}$}
\FAProp(0.,10.)(11.,10.)(0.,){/ScalarDash}{0}
\FALabel(5.5,9.18)[t]{$\Phi$}
\FAProp(20.,15.)(11.,10.)(0.,){/Straight}{-1}
\FALabel(15.2273,13.3749)[br]{$f_1$}
\FAProp(20.,5.)(17.3,6.5)(0.,){/Straight}{1}
\FALabel(18.9227,6.62494)[bl]{$f_2$}
\FAProp(11.,10.)(13.7,8.5)(0.,){/Straight}{-1}
\FALabel(12.0773,8.37506)[tr]{$f_1$}
\FAProp(17.3,6.5)(13.7,8.5)(-0.8,){/Straight}{1}
\FALabel(14.4273,5.18506)[tr]{$f^\prime_i$}
\FAProp(17.3,6.5)(13.7,8.5)(0.8,){/Sine}{-1}
\FALabel(16.5727,9.81494)[bl]{$W$}
\FAVert(11.,10.){0}
\FAVert(17.3,6.5){0}
\FAVert(13.7,8.5){0}

\FADiagram{d$_{9}$}
\FAProp(0.,10.)(11.,10.)(0.,){/ScalarDash}{0}
\FALabel(5.5,9.18)[t]{$\Phi$}
\FAProp(20.,15.)(17.3,13.5)(0.,){/Straight}{-1}
\FALabel(18.3773,15.1249)[br]{$f_1$}
\FAProp(20.,5.)(11.,10.)(0.,){/Straight}{1}
\FALabel(15.2273,6.62506)[tr]{$f_2$}
\FAProp(11.,10.)(13.7,11.5)(0.,){/Straight}{1}
\FALabel(12.0773,11.6249)[br]{$f_2$}
\FAProp(17.3,13.5)(13.7,11.5)(-0.8,){/Straight}{-1}
\FALabel(16.5727,10.1851)[tl]{$f^\prime_i$}
\FAProp(17.3,13.5)(13.7,11.5)(0.8,){/ScalarDash}{1}
\FALabel(14.4273,14.8149)[br]{$G$}
\FAVert(11.,10.){0}
\FAVert(17.3,13.5){0}
\FAVert(13.7,11.5){0}

\FADiagram{d$_{10}$}
\FAProp(0.,10.)(11.,10.)(0.,){/ScalarDash}{0}
\FALabel(5.5,9.18)[t]{$\Phi$}
\FAProp(20.,15.)(17.3,13.5)(0.,){/Straight}{-1}
\FALabel(18.3773,15.1249)[br]{$f_1$}
\FAProp(20.,5.)(11.,10.)(0.,){/Straight}{1}
\FALabel(15.2273,6.62506)[tr]{$f_2$}
\FAProp(11.,10.)(13.7,11.5)(0.,){/Straight}{1}
\FALabel(12.0773,11.6249)[br]{$f_2$}
\FAProp(17.3,13.5)(13.7,11.5)(-0.8,){/Straight}{-1}
\FALabel(16.5727,10.1851)[tl]{$f^\prime_i$}
\FAProp(17.3,13.5)(13.7,11.5)(0.8,){/Sine}{1}
\FALabel(14.4273,14.8149)[br]{$W$}
\FAVert(11.,10.){0}
\FAVert(17.3,13.5){0}
\FAVert(13.7,11.5){0}

\FADiagram{d$_{11}$}
\FAProp(0.,10.)(6.5,10.)(0.,){/ScalarDash}{0}
\FALabel(3.25,9.18)[t]{$\Phi$}
\FAProp(20.,15.)(13.,14.)(0.,){/Straight}{-1}
\FALabel(16.2808,15.5544)[b]{$f_1$}
\FAProp(20.,5.)(13.,6.)(0.,){/Straight}{1}
\FALabel(16.2808,4.44558)[t]{$f_2$}
\FAProp(6.5,10.)(13.,14.)(0.,){/Straight}{1}
\FALabel(9.20801,13.1807)[br]{$f^\prime_i$}
\FAProp(6.5,10.)(13.,6.)(0.,){/Straight}{-1}
\FALabel(9.20801,6.81927)[tr]{$f^\prime_i$}
\FAProp(13.,14.)(13.,6.)(0.,){/ScalarDash}{1}
\FALabel(14.274,10.)[l]{$H$}
\FAVert(6.5,10.){0}
\FAVert(13.,14.){0}
\FAVert(13.,6.){0}

\FADiagram{d$_{12}$}
\FAProp(0.,10.)(6.5,10.)(0.,){/ScalarDash}{0}
\FALabel(3.25,9.18)[t]{$\Phi$}
\FAProp(20.,15.)(13.,14.)(0.,){/Straight}{-1}
\FALabel(16.2808,15.5544)[b]{$f_1$}
\FAProp(20.,5.)(13.,6.)(0.,){/Straight}{1}
\FALabel(16.2808,4.44558)[t]{$f_2$}
\FAProp(6.5,10.)(13.,14.)(0.,){/ScalarDash}{-1}
\FALabel(9.20801,13.1807)[br]{$H$}
\FAProp(6.5,10.)(13.,6.)(0.,){/ScalarDash}{1}
\FALabel(9.20801,6.81927)[tr]{$H$}
\FAProp(13.,14.)(13.,6.)(0.,){/Straight}{-1}
\FALabel(14.274,10.)[l]{$f^\prime_i$}
\FAVert(6.5,10.){0}
\FAVert(13.,14.){0}
\FAVert(13.,6.){0}

\FADiagram{d$_{13}$}
\FAProp(0.,10.)(6.5,10.)(0.,){/ScalarDash}{0}
\FALabel(3.25,9.18)[t]{$\Phi$}
\FAProp(20.,15.)(13.,14.)(0.,){/Straight}{-1}
\FALabel(16.2808,15.5544)[b]{$f_1$}
\FAProp(20.,5.)(13.,6.)(0.,){/Straight}{1}
\FALabel(16.2808,4.44558)[t]{$f_2$}
\FAProp(6.5,10.)(13.,14.)(0.,){/ScalarDash}{-1}
\FALabel(9.20801,13.1807)[br]{$H$}
\FAProp(6.5,10.)(13.,6.)(0.,){/ScalarDash}{1}
\FALabel(9.20801,6.81927)[tr]{$G$}
\FAProp(13.,14.)(13.,6.)(0.,){/Straight}{-1}
\FALabel(14.274,10.)[l]{$f^\prime_i$}
\FAVert(6.5,10.){0}
\FAVert(13.,14.){0}
\FAVert(13.,6.){0}

\FADiagram{d$_{14}$}
\FAProp(0.,10.)(6.5,10.)(0.,){/ScalarDash}{0}
\FALabel(3.25,9.18)[t]{$\Phi$}
\FAProp(20.,15.)(13.,14.)(0.,){/Straight}{-1}
\FALabel(16.2808,15.5544)[b]{$f_1$}
\FAProp(20.,5.)(13.,6.)(0.,){/Straight}{1}
\FALabel(16.2808,4.44558)[t]{$f_2$}
\FAProp(6.5,10.)(13.,14.)(0.,){/ScalarDash}{-1}
\FALabel(9.20801,13.1807)[br]{$G$}
\FAProp(6.5,10.)(13.,6.)(0.,){/ScalarDash}{1}
\FALabel(9.20801,6.81927)[tr]{$H$}
\FAProp(13.,14.)(13.,6.)(0.,){/Straight}{-1}
\FALabel(14.274,10.)[l]{$f^\prime_i$}
\FAVert(6.5,10.){0}
\FAVert(13.,14.){0}
\FAVert(13.,6.){0}

\FADiagram{d$_{15}$}
\FAProp(0.,10.)(6.5,10.)(0.,){/ScalarDash}{0}
\FALabel(3.25,9.18)[t]{$\Phi$}
\FAProp(20.,15.)(13.,14.)(0.,){/Straight}{-1}
\FALabel(16.2808,15.5544)[b]{$f_1$}
\FAProp(20.,5.)(13.,6.)(0.,){/Straight}{1}
\FALabel(16.2808,4.44558)[t]{$f_2$}
\FAProp(6.5,10.)(13.,14.)(0.,){/ScalarDash}{-1}
\FALabel(9.20801,13.1807)[br]{$H$}
\FAProp(6.5,10.)(13.,6.)(0.,){/Sine}{1}
\FALabel(9.20801,6.81927)[tr]{$W$}
\FAProp(13.,14.)(13.,6.)(0.,){/Straight}{-1}
\FALabel(14.274,10.)[l]{$f^\prime_i$}
\FAVert(6.5,10.){0}
\FAVert(13.,14.){0}
\FAVert(13.,6.){0}

\FADiagram{d$_{16}$}
\FAProp(0.,10.)(6.5,10.)(0.,){/ScalarDash}{0}
\FALabel(3.25,9.18)[t]{$\Phi$}
\FAProp(20.,15.)(13.,14.)(0.,){/Straight}{-1}
\FALabel(16.2808,15.5544)[b]{$f_1$}
\FAProp(20.,5.)(13.,6.)(0.,){/Straight}{1}
\FALabel(16.2808,4.44558)[t]{$f_2$}
\FAProp(6.5,10.)(13.,14.)(0.,){/Sine}{-1}
\FALabel(9.20801,13.1807)[br]{$W$}
\FAProp(6.5,10.)(13.,6.)(0.,){/ScalarDash}{1}
\FALabel(9.20801,6.81927)[tr]{$H$}
\FAProp(13.,14.)(13.,6.)(0.,){/Straight}{-1}
\FALabel(14.274,10.)[l]{$f^\prime_i$}
\FAVert(6.5,10.){0}
\FAVert(13.,14.){0}
\FAVert(13.,6.){0}

\FADiagram{d$_{17}$}
\FAProp(0.,10.)(11.,10.)(0.,){/ScalarDash}{0}
\FALabel(5.5,9.18)[t]{$\Phi$}
\FAProp(20.,15.)(11.,10.)(0.,){/Straight}{-1}
\FALabel(15.2273,13.3749)[br]{$f_1$}
\FAProp(20.,5.)(17.3,6.5)(0.,){/Straight}{1}
\FALabel(18.9227,6.62494)[bl]{$f_2$}
\FAProp(11.,10.)(13.7,8.5)(0.,){/Straight}{-1}
\FALabel(12.0773,8.37506)[tr]{$f_1$}
\FAProp(17.3,6.5)(13.7,8.5)(-0.8,){/Straight}{1}
\FALabel(14.4273,5.18506)[tr]{$f^\prime_i$}
\FAProp(17.3,6.5)(13.7,8.5)(0.8,){/ScalarDash}{-1}
\FALabel(16.5727,9.81494)[bl]{$H$}
\FAVert(11.,10.){0}
\FAVert(17.3,6.5){0}
\FAVert(13.7,8.5){0}

\FADiagram{d$_{18}$}
\FAProp(0.,10.)(11.,10.)(0.,){/ScalarDash}{0}
\FALabel(5.5,9.18)[t]{$\Phi$}
\FAProp(20.,15.)(17.3,13.5)(0.,){/Straight}{-1}
\FALabel(18.3773,15.1249)[br]{$f_1$}
\FAProp(20.,5.)(11.,10.)(0.,){/Straight}{1}
\FALabel(15.2273,6.62506)[tr]{$f_2$}
\FAProp(11.,10.)(13.7,11.5)(0.,){/Straight}{1}
\FALabel(12.0773,11.6249)[br]{$f_2$}
\FAProp(17.3,13.5)(13.7,11.5)(-0.8,){/Straight}{-1}
\FALabel(16.5727,10.1851)[tl]{$f^\prime_i$}
\FAProp(17.3,13.5)(13.7,11.5)(0.8,){/ScalarDash}{1}
\FALabel(14.4273,14.8149)[br]{$H$}
\FAVert(11.,10.){0}
\FAVert(17.3,13.5){0}
\FAVert(13.7,11.5){0}

\FADiagram{}

\FADiagram{}

\FADiagram{}

\FADiagram{}

\FADiagram{}

\FADiagram{}

\FADiagram{}

\FADiagram{}

\FADiagram{}

\FADiagram{}

\FADiagram{}

\FADiagram{}

\FADiagram{}

\FADiagram{}
\end{feynartspicture}